\documentclass[showpacs,preprintnumbers,amsmath,amssymb,superscriptaddress]{revtex4}

\usepackage{graphicx}% Include figure files
\usepackage{dcolumn}% Align table columns on decimal point
\usepackage{bm}% bold math
\usepackage{graphics}
\input{epsf}

\newcommand{\beq}{\begin{equation}}
\newcommand{\eeq}{\end{equation}}
\def\la{\hbox{\raise.35ex\rlap{$<$}\lower.6ex\hbox{$\sim$}\ }}
\def\ga{\hbox{\raise.35ex\rlap{$>$}\lower.6ex\hbox{$\sim$}\ }}

\def\beq{\begin{equation}}
\def\eeq{\end{equation}}
\def\beqa{\begin{eqnarray}}
\def\eeqa{\end{eqnarray}}
\def\bseq{\begin{subequations}}
\def\eseq{\end{subequations}}

\def\order#1{{\cal O}\left({#1}\right)}
\newcommand{\sfrac}[2]{\mbox{$\frac{#1}{#2}$}}
\begin{document}

\preprint{APS/123-QED}

\title{On the nonlinear saturation of the magnetorotational instability
near threshold\\ in a thin-gap Taylor-Couette setup }

\author{O.M. Umurhan}
\affiliation{Department of Physics, Technion-Israel Institute of
Technology, 32000 Haifa, Israel}
\email[]{mumurhan@physics.technion.ac.il}
 \affiliation{Department of Geophysics and Planetary Sciences, Tel--Aviv
University, Israel} \affiliation{Department of Astronomy, City
College San Francisco, SF, CA 94112, USA}
\author{O. Regev}
\affiliation{Department of Physics, Technion-Israel Institute of
Technology, 32000 Haifa, Israel}
\email[]{regev@physics.technion.ac.il}
\affiliation{Department of
Astronomy, Columbia University, NY, NY 10027 USA}
\author{K. Menou}
\affiliation{Department of Astronomy,
Columbia University, NY, NY 10027 USA} \
\date{\today}% It is always \today, today,

\begin{abstract}
We study the saturation near threshold of the axisymmetric
magnetorotational instability (MRI) of a viscous, resistive,
incompressible fluid in a thin-gap Taylor-Couette configuration. A
vertical magnetic field, Keplerian shear and no-slip, conducting
radial boundary conditions are adopted. The weakly non-linear theory
leads to a real Ginzburg-Landau equation for the disturbance
amplitude, like in our previous idealized analysis.  For small
magnetic Prandtl number (${\cal P}_{\rm m} \ll 1$), the saturation
amplitude scales as ${\cal P}_{\rm m}^{2/3}$
while the magnitude of angular momentum transport scales
as ${\cal P}_{\rm m}^{4/3}$. The difference with the previous scalings
($\propto {\cal P}_{\rm m}^{1/2}$ and ${\cal P}_{\rm m}$ respectively)
is attributed to the emergence of radial
boundary layers. Away from those, steady-state non-linear saturation
is achieved through a modest reduction in the destabilizing
shear. These results will be useful to understand MRI laboratory
experiments and associated numerical simulations.
\end{abstract}

\pacs{Valid PACS appear here}% PACS, the Physics and Astronomy
\maketitle
\section{Introduction}

The magneto-rotational instability (MRI) is a linear instability
known to occur in rotating hydromagnetic shear flows when the
angular velocity decreases with distance from the rotation axis,
i.e. $\partial_R (\Omega^2)<0$. Although it had been known for
almost half a century \cite{veli,chandra1, chandra2}, the MRI
acquired a renewed interest only after the influential work of
Balbus \& Hawley \cite{BH91}, who have shown, by means of linear
stability analysis and numerical simulations, its viability in
conditions locally approximating astrophysical accretion disks.
Subsequent investigations of this kind (see the reviews by Balbus
\& Hawley \cite{BHrev, B03} and references therein) have quite
convincingly demonstrated that this instability can drive
magnetohydrodynamical (MHD) turbulence in a variety of conditions,
appropriate to accretion disks and more general settings as well.
Within the framework of a magnetic Taylor-Couette configuration,
which is relevant for the present work, the parameter dependencies
(magnetic Reynolds and Prandtl numbers) of the marginal (linear)
instability threshold has been previously considered
\cite{RS02,WB02}. It was found, among other things, that the
critical magnetic Reynolds number does not scale with the magnetic
Prandtl number, for small values of the latter.  This result
carries over into the weakly nonlinear theory presented here.

Accretion disks are important and ubiquitous astrophysical objects and
are thought to power as diverse systems as young stellar objects,
close binary systems and active galactic nuclei.  Accretion disks are
flattened, high specific angular momentum (with essentially a
Keplerian distribution) masses of gas, through which matter accretes
onto a central object. An efficient dissipation and transport of
angular momentum mechanism is needed in order to allow accretion and
reconcile theoretical models with observations.  Since the typical
hydrodynamical Reynolds numbers (${\cal R}$) in these astrophysical
flows are enormous, it has been recognized at the outset, when
accretion disks were theoretically proposed \cite{SS, LBP}, that some
anomalous, enhanced (conceivably turbulent) dissipation and transport
must be invoked.  Keplerian rotating flows are (according to the
Rayleigh and other criteria) linearly stable and thus astrophysical
disk turbulence can not originate from a linear instability of the
kind known (and well studied) in Taylor-Couette hydrodynamical
flows.
%Several ideas on how purely hydrodynamical processes may
%account for anomalous angular momentum transport in accretion disks
%have been put forward
%\cite{dubrulle,jpz,petros,longaretti1,chagelishvili03,kb,
%yecko04,ur04,dubrulle2,Narayan2,unrs,banibrata}, but they still remain
%quite controversial \cite{balbusfight,gb-nonpub,longaretti2,ogilvie,
%matt,bh06,shenstone,nature} and we shall not deal with this
%problem here.

The physics of the non-linear development of the MRI, its saturation
and the nature of the resulting angular momentum transport are quite
complicated.  Almost all of our present knowledge on this subject
comes from numerical simulations, carried out by several groups (see,
e.g., \cite{BH91, balbusfight, BHrev, B03} and references
therein). These finite-difference simulations, even though intended
for the study of the MRI in its astrophysical setting, were actually
local, i.e. done for a small portion of an accretion disk, in what is
known as the {\it shearing box} or {\it sheet} (hereafter, SB)
formulation \cite{glb}, \cite{BH91} (see the Appendix of \cite{ur04}
for a formal account on this approximation).  Although a lot has been
learned from these simulations, the intricate processes at work are
not yet fully understood and some basic physical questions remain open
(see, e.g., \cite{brand}). As a result, there has recently been a
growing interest to observe the instability in the laboratory, where
various physical aspects can be unraveled in a controlled way.  A
number of groups have indeed embarked on such experimental projects,
in several setups, often accompanying them by appropriate numerical
calculations (e.g.,
\cite{jgk,colgate,sisan,hollerbach,liu_et_al,helical}, and references
therein).

In comparison to the large extent of numerical and experimental work
on the MRI's nonlinear development, there have only been very few
reports on analytical and semi-analytical studies on this
subject. This fact seems surprising, because a very large body of
work, utilizing various asymptotic approaches, has been done for other
important fluid instabilities (for reviews see, e.g.,
\cite{manneville1,cross,manneville2,regev,pismen}). We are aware of
only two asymptotic studies of this kind in the MRI context:
\begin{itemize}
\item
In the first one \cite{knobloch}, Knobloch \& Julien investigated the
saturation of the MRI in the strongly nonlinear (far from instability
threshold) regime. They utilized the so-called channel modes (radially
independent axisymmetric linear modes, which also happen to be exact
solutions of the nonlinear problem in the SB formulation
\cite{BHrev},\cite{GX}). They performed an asymptotic calculation, in
which the evolution of channel modes is followed into the nonlinear
regime by gently tuning the system out of the developed
short-wavelength channel mode configuration (and under a specific
regime of system's parameters). This work shows that nonlinearities
saturate the system in such a way that the momentum transport scales
as $({\cal R R}_m)^{-1}$, where ${\cal R}$ and ${\cal R}_m$ are the
hydrodynamic and magnetic Reynolds numbers, respectively (see their
Eq. 4.22).  The results further indicate that, by modifying the
underlying shear (the ``source" of instability), the system saturates
while approaching solid body rotation.
\item
In the second study \cite{omr}, hereafter UMR06, we have employed a
more traditional approach - weakly nonlinear asymptotics close to the
instability threshold.  The problem we considered differed from
previous studies in that we considered the dynamics to be restricted
to a narrow (in its radial extent) channel.  Our original intent was
to understand the MRI under a more controlled setting - one in which
the channel modes are filtered out by the imposition of no normal-flow
conditions at the inner and outer boundaries of the channel.  Under
these conditions, arguably more appropriate to capture the physics of
experimental setups, the MRI unstable mode transits into instability
in a way analogous to that of Rayleigh-B\'{e}nard convection.  An
idealization, involving a hybrid free-slip/no-slip and
conducting/insulating boundary conditions, atop the no-normal flow
conditions mentioned above, allows for transparent analytical
evaluations of the derived necessary quantities (similar idealizations
have sometimes been used in other studies \cite{liu_et_al}) of the
problem. The similarity of this formulation to other extensively
studied hydrodynamical instability problems led us to the application
of weakly nonlinear asymptotic techniques to examine the system's
transition into the nonlinear realm, as well as to comparison of the
results to specially-designed numerical simulations.  We found that,
as the system is gently tuned into instability (through a suitably
defined non-dimensional parameter $\epsilon$), a saturated
pattern-state emerges with an amplitude of the most unstable mode
evolving according to the real Ginzburg-Landau equation (GLE),
\begin{equation}
\partial_{_T} A = \lambda A + D \partial^2_{_Z} A - \alpha |A|^2A = 0,
\label{our_rGLE}
\end{equation}
where $T$ and $Z$ are suitably ``stretched" time and vertical
coordinates and the coefficients of the equation are all real and
computable from the parameters of the physical problem. In particular,
the coefficient $\alpha$ was found to scale as ${\cal P}_m^{-1}$,
where ${\cal P}_m$ is the {\em magnetic} Prandtl number, defined by
the ratio ${\cal R}_m/{\cal R}$.  It means that the amplitude achieved
by the system in the saturated state scales as $\epsilon \sqrt {{\cal
P}_m}$ and correspondingly, the overall angular momentum transport as
$\epsilon^2 {\cal P}_m$. For ${\cal R}_m$ {\em fixed} this transport
would scale like $1/{\cal R}$ and this formulation is useful when the
resistivity of the medium is set by its physical state (i.e. degree of
ionization) and one wishes to estimate the effect of decreasing
effective viscosity (resulting, e.g. from the inaccuracy of the
numerical scheme in a simulation). These analytical scalings were
found in the limit ${\cal P}_m \ll 1$, while for larger values of
${\cal P}_m$ similar trends may be expected but the coefficients have
to be evaluated numerically. We have conjectured that for
self-consistent boundary conditions the above general qualitative
behavior should hold as well, with perhaps some change in the relevant
power of ${\cal P}_m$ in the scalings. Our asymptotic analysis was
accompanied by fully numerical spectral calculations of the original
SB equations with similarly idealized boundary conditions. The
analytical and numerical scalings were found to agree quite well.
\end{itemize}

In this paper we present a study of the MRI as developing in a model
representing the thin-gap limit of a magnetic Taylor-Couette
(hereafter mTC) configuration, in which an incompressible axisymmetric
rotating flow is subject to an external vertical magnetic field.  This
will permit a quantitative examination of the effect of the boundary
conditions on the results reported in UMR06 and confirm the conjecture
on the general qualitative behavior.

The fundamental equations of motion are the same as those assumed in
previous studies of the MRI (e.g. \cite{BH91}) save for the inclusion
of non-ideal effects, namely resistivity and viscosity.  Solutions to
these equations are sought, subject to realistic boundary conditions
at the system walls, namely that of no-flow and conducting conditions.
For the vertical boundary conditions we assume periodicity for the
sake of simplicity and transparency. After presenting, in Section
\ref{equationes}, the relevant approximations, definitions and
equations, we perform, in Section \ref{linear_theory}, a linear
eigenmode analysis.  We identify the most unstable mode as a function
of the non-dimensional parameters of the system - of which there are
five: the Cowling number ${\cal C}$, the magnetic Prandtl number
${\cal P}_m$, the magnetic Reynolds number ${\cal R}_m$, and shear
index $q$ (see below).  We demonstrate next that this system has a
transition into instability which is similar in some important aspects
to that in Rayleigh-B\'{e}nard convection
\cite{manneville1,cross,manneville2,regev}.  We also identify the
presence of a neutral, spatially constant mode representing the hand
of a constant azimuthal field.  \par In Section \ref{weakly_nonlinear}
we perform a weakly nonlinear asymptotic analysis by tuning the system
away from the conditions of marginality. In this case this is done by
ratcheting the background magnetic field downward from the marginal
state with the magnitude of the departure from that state measured by
the small parameter $\epsilon^2$.  The full calculation, detailed in
Appendices B-D, reveals that the envelope (of the marginally unstable
modes) evolution is governed by two uncoupled partial differential
equations: one represents the leading MRI mode and evolves according
to the real GLE and the other equation, representing the evolution of
the uniform azimuthal field, is a standard diffusion equation.  The
saturated amplitude of the leading MRI mode is demonstrated, in the
${\cal P}_m \ll 1$ limit, to scale as $\epsilon{\cal P}_m^{2/3}$ and
is shown to be affected by the boundary layers appearing at the system
walls. The main physical factor contributing to saturation is
identified as coming from the second order (in $\epsilon$) correction
to the azimuthal velocity perturbation in the limit ${\cal P}_m \ll
1$.  This, in turn, affects the shear profile so as to stabilize the
new steady configuration. We also find that the average total angular
momentum transport implied under these conditions scales as
$\epsilon^2 {\cal P}_m^{4/3}$ for ${\cal P}_m \ll 1$, or as
$\epsilon^2 {\cal R}^{-4/3}$ for ${\cal R}_m$ {\em fixed} (and of
$\order{1}$). These results are in accord with our conjecture and
expectations given in UMR06.

In the last Section we discuss the implications of our work and how it
should be perceived as a part of the ongoing research efforts on
various aspects of the MRI. We also provide some heuristic arguments
to help understand the results.  Finally, we end with a short outline
of possible directions for future work of this kind.

\section{Assumptions, definitions and equations}
\label{equationes}
The hydromagnetic equations in cylindrical coordinates \cite{chandra2}
are applied to the neighborhood of a representative radial point
($r_0$) in the system, using the above mentioned shearing box (SB)
approximation. The SB is applied here to the thin-gap limit of a
Taylor-Couette setup with an imposed background vertical magnetic
field.  We begin by considering a steady base flow with only a
constant vertical magnetic field, ${\bf B} = B_0 {\bf \hat z}$, and a
velocity of the form ${\bf V}=U(x){\bf {\hat y}} $. In this base state
the velocity has a linear shear profile $U(x)=-q\Omega_0 x$,
representing an azimuthal flow about a point $r_0$, that rotates with
a rate $\Omega_0$, defined from the differential rotation law
$\Omega(r) \propto \Omega_0 (r/r_0)^{-q}$.  The total pressure in the
base state (divided by the constant density),
\[ \Pi \equiv \frac{1}{\rho_0}\left(P+\frac{B_0^2}{8\pi}\right),\]
is a constant and thus its gradient is zero.

This base flow is disturbed by 3-D perturbations on the magnetic field
${\bf b}= (b_x,b_y,b_z)$, as well as on the velocity - ${\bf
u}=(u_x,u_y,u_z)$, and on the total pressure - $\varpi$.  We consider
only axisymmetric disturbances, i.e. perturbations with structure only
in the $x$ and $z$ directions.  This results, after
non-dimensionalization, in the following set of {\em non-linear}
equations: \beqa \frac{d{\bf u}}{dt} -2{ {\Omega_0 \bf{ \hat
z}}\times{\bf u}} -{q\Omega_0 u_x {\bf{\hat y}}} - {\cal C} {\bf
b}\cdot\nabla{\bf b} -{\cal C}B_0\partial_z {\bf b} &=& -\nabla\varpi
+\frac{1}{{\cal R}}\nabla^2 {\bf u}, \\ \frac{d{\bf b}}{dt} - {\bf
b}\cdot\nabla{\bf u} + {q\Omega_0 b_x {\bf{\hat y}}} -B_0\partial_z
{\bf u} &=& \frac{1}{{\cal R}_{{ m}}}\nabla^2 {\bf b}, \eeqa together
with an incompressibility condition and the solenoidal magnetic field
constraint \beq \nabla\cdot{\bf u} \equiv \partial_x u_x + \partial_z
u_z = 0, \qquad \nabla\cdot{\bf b} \equiv \partial_x b_x + \partial_z
b_z = 0.
\label{incompressible_conditions}
\eeq The Cartesian coordinates $x,y,z$ represent here the radial
(shear-wise), azimuthal (stream-wise) and vertical directions
respectively and since axisymmetry is assumed $\nabla \equiv {\bf \hat
x}\partial_x+ {\bf \hat z}\partial_z$ and the Laplacian is $\nabla^2
\equiv \partial_x^2 + \partial_z^2$.  Lengths have been
non-dimensionalized by $L$ (the shearing-box size), time $t$ by the
local rotation rate $\tilde\Omega_0$ (tildes denote here dimensional
quantities).  Because the dimensional rotation rate of the box (about
the central object) is ${\bf \tilde \Omega_0} = \tilde \Omega_0{\bf
\hat z}$, the non-dimensional quantity $\Omega_0$ is formally
equivalent to $1$, but we keep it to flag the Coriolis terms.
Velocities have been scaled by $\tilde \Omega_0 L$ and the magnetic
field by the value of the background vertical field $\tilde B_0$.
Thus the non-dimensional constant background field $B_0 \equiv 1$, but
again, we leave it in the equation set for later convenience (see
below). The hydrodynamic pressure is scaled by $\tilde\rho_0 L^2
\tilde \Omega_0^2$ and the magnetic one by $\tilde B_0^2/(8 \pi)$.
The non-dimensional perturbation $\varpi$ of the total pressure
divided by the density (which is equal to 1 in non-dimensional units),
which survives the spatial derivatives, is thus given by \beq \varpi =
p + {\cal C} \sfrac{1}{2} |{\bf b}|^2, \eeq where $p$ is the
hydrodynamic pressure perturbation.  \par The non-dimensional
parameter \beq {\cal C} \equiv \frac{\tilde B_0^2}{4\pi\tilde
\rho_0\tilde\Omega_0^2L^2} = \frac{\tilde V_A^2}{\tilde V^2}
\label{alfven_number} \eeq
is the {\em Cowling number}, measuring the relative importance of the
magnetic pressure to the hydrodynamical one. It is equal to the
inverse square of the typical Alfv\'en number ($\tilde V_A$ is the
typical Alfv\'en speed).  The Cowling number appears in the non-linear
equations, together with the two {\em Reynolds numbers}
\beq
{\cal R} \equiv \frac{\tilde\Omega_0 L^2}{\tilde \nu}, \qquad
{\cal R}_m \equiv \frac{\tilde\Omega_0 L^2}{\tilde \eta},
\eeq
where $\tilde \nu$ and $\tilde \eta$ are, respectively, the
microscopic viscosity and magnetic resistivity of the fluid.  We shall
also see that the {\em magnetic Prandtl number}, given as ${\cal P}_m
\equiv {\cal R}_{ m}/{\cal R}$, plays an important role in the
nonlinear evolution of this system.

We rewrite now the equations of motion in terms of more convenient
dependent variables:
\beqa
\partial_t\nabla^2\Psi + N_{_\Psi}&=&
{{\cal R}}^{^{-1}}
\nabla^4\Psi
+ 2\Omega_0\partial_z u_y +
{\cal C}B_0\partial_z\nabla^2\Phi
\label{psi_equation} \\
\partial_t u_y  + N_{_u} &=&
{{\cal R}}^{^{-1}}\nabla^2 u_y
- \Omega_0(2-q)\partial_z\Psi +
{\cal C}B_0\partial_z b_y
\\
\partial_t\Phi + N_{_\Phi} &=&
{{\cal R}_m}^{^{-1}}\nabla^2 \Phi + B_0\partial_z\Psi\label{Phi_equation}
 \\
\partial_t b_y + N_{_b}&=&{{\cal R}_m}^{^{-1}}\nabla^2
b_y
+B_0\partial_z u_y
-\underline{q\Omega_0\partial_z\Phi}
,
\label{by_equation}
\eeqa

Because the flow is incompressible and $y$-independent, the radial and
vertical velocities are expressed in terms of the {\em
streamfunction}, $\Psi$, that is, $(u{_x},u{_z}) =
(\partial_z\Psi,-\partial_x \Psi)$. Also, since the magnetic field is
source free, we similarly express its vertical and radial components
in terms of the {\em flux function}, $\Phi$, that is, $(b{_x},b{_z})
=(\partial_z\Phi,-\partial_x \Phi)$. Note that (\ref{Phi_equation})
combines information about the radial and vertical magnetic fields in
terms of the flux and streamfunctions (e.g. \cite{liu_et_al}).  In
this formulation the nonlinear advection and tension terms are \beqa
N_{_\Psi} \equiv&& J(\Psi,\nabla^2\Psi) - {\cal
C}J(\Phi,\nabla^2\Phi), \quad N_{_u} \equiv J(\Psi,u_y) - {\cal
C}J(\Phi,b_y),\nonumber\\ \quad N_{_\Phi} \equiv&& J(\Psi,\Phi), \quad
N_{_b} \equiv J(\Psi,b_y) - J(\Phi,u_y), \eeqa in which the Jacobian
is defined as $J(f,g) \equiv \partial_z f\, \partial_x g -\partial_x
f\, \partial_z g$.  The underlined term in (\ref{by_equation}),
representing the transport of the perturbed radial magnetic field by
the background shear flow, is instrumental for the occurrence of the
MRI in this system.

The boundary conditions are periodic on the vertical boundaries of the
domain and we require also that the flow be no-slip at the inner and
outer boundaries.  This means that $\bf u = 0$ at $x=\pm 1$, i.e.
\beq
u_y = 0,~~\partial_z \Psi =0,~~ \partial_x \Psi = 0, \qquad {\rm at} \ \ x = \pm 1.
\label{u_bc}
\eeq
Regarding the boundary conditions on the magnetic field disturbances,
we posit conditions (only 2 are needed) that are consistent with the
inner and outer walls being conducting, $b_x=0$ and $\partial_x
b_y=0$ at $x = \pm 1$, i.e.  \beq \partial_z \Phi = 0,~~\partial_x b_y
= 0, \qquad {\rm at} \ \ x = \pm 1.
\label{b_bc}
\eeq
Note that these boundary conditions are more physically consistent
than the ones we have used in UMR06, however they will call for a
numerical evaluation of the eigenfunctions and the coefficients for
the asymptotic analysis that result from them.

\par
Finally, we point out that there exists an energy theorem for the above
dynamical equations. Defining the total energy (per unit length in the
azimuthal direction) of the disturbances in the domain as $E
\equiv \sfrac{1}{2}\int{\left({\bf u}^2 + {\cal C}{\bf
b}^2\right)dxdz}$, we get, after the usual integration procedures
and application of boundary conditions,

\beqa
 \frac{dE}{dt} = q\Omega_0 \int{{\mathbb T} dx dz}
-\frac{1}{\cal R}\int {\left(|\nabla u_x|^2 + |\nabla u_y|^2 +
|\nabla u_z|^2 \right)dxdz}
-\frac{{\cal C}}{{\cal R}_m} \int{\left(|\nabla b_x|^2 + |\nabla b_y|^2 + |\nabla b_z|^2
\right)dxdz},
\label{energy_theorem}
\eeqa
where
\[
{\mathbb T} ={\mathbb T}_{\rm R}+{\mathbb T}_{\rm M},\qquad
{\mathbb T}_{\rm R} \equiv
u_x u_y, \quad {\mathbb T}_{\rm M} \equiv -{\cal C} b_x b_y.
\]
${\mathbb T}_{\rm R}$ and ${\mathbb T}_{\rm M}$ are the Reynolds
(hydrodynamic) and Maxwell stresses, capturing the velocity and
magnetic field disturbance correlations, respectively. Statement
(\ref{energy_theorem}) is analogous to the Reynolds-Orr relation in
hydrodynamics (for which ${\mathbb T}={\mathbb T}_{\rm R}$ and
${\mathbb T}_{\rm R}=0$).  The total stress ${\mathbb T}$ will be used
in the asymptotic theory we develop here as the dominant expression
for the evaluation of transport, occurring during the weakly nonlinear
evolution of the system. The full RHS of (\ref{energy_theorem}),
including the two dissipative terms, should obviously vanish when a
saturated, steady state is reached. We discuss this in more detail in
Section V.

\begin{figure}
\begin{center}
\leavevmode \epsfysize=5.0
cm
\epsfbox{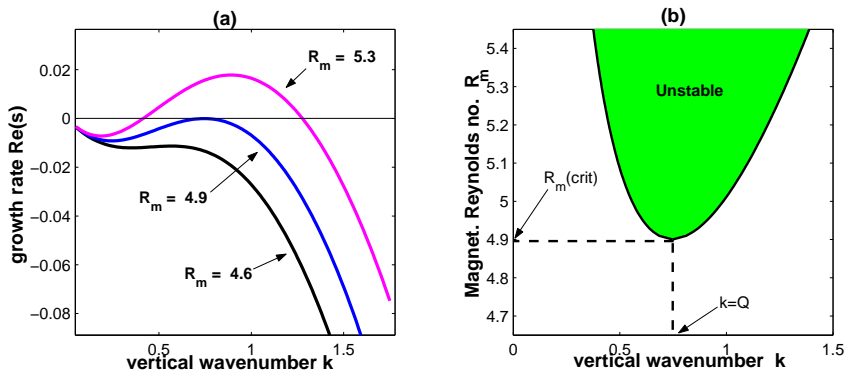}
\end{center}
\caption{{\small (Color online)
Summary of linear theory.  This example is for ${\cal C} = 0.08$,
${\cal P}_m = 0.001$, $q = 3/2$, and the fundamental mode.
(a) Growth rates, $Re(s)$, as a function
of wavenumber $k$ for three values of ${\cal R}_{m}$.  (b) Solid line
depicts those values of ${\cal R}_{m}$ and $k$ where $Re(s)=0$.  The
shaded region shows unstable modes.  The locations of $k=k_{crit}\equiv Q$ and
${\cal R}_m= {\cal R}_{m}({\rm crit}) \equiv  R_m$ are shown.
}}
\label{linear_plottery_Pm001}
\end{figure}

\section{Linear Theory}\label{linear_theory}
Linearization of (\ref{psi_equation}-\ref{by_equation}) yields the following
equation
\beq
\partial_t {\cal D} {{\mathbf V}_1} = {\cal L} { \mathbf V}_1,
\label{linear_operator}
\eeq
in which all the small perturbations are lumped in the vector
${\mathbf V}_1 = \left(\Psi_{_1}(x),u_{_1}(x),\Phi_{_1}(x),
b_{_1}(x)\right)^{{\mathbf T}} e^{ikz + s t} + {\rm c.c.}$, with $k$
being the vertical wave-number and $s$ the temporal eigenvalue. The
spatial differential operators $\cal D$ and $\cal L$ (appropriately
written in the form of $4\times4$ matrices) are explicitly given in
%(\ref{L_Definition}-\ref{matrix_definitions})
(B1-B3) of Appendix B.  As long as $k\neq 0$ the boundary conditions
on the functions of ${ \mathbf V}_1$ become (see
eqs. \ref{u_bc}-\ref{b_bc})
\beq
\Psi_{_1} = D_x \Psi_{_1} = u_{_1} = \Phi_{_1} = D_x b_{_1} = 0, \qquad {\rm at}
\ \ x = \pm 1,
\label{bcc}
\eeq
where $D_x \equiv d/dx$.

In principle, equations (\ref{linear_operator}) can be set up and
solved analytically however the resulting expressions are far too
cumbersome to be conveniently manipulated.  It is much easier to solve
this set numerically, using a Chebyshev collocation technique. Each
function is approximated using typically between 30 and 60 grid points
on a Chebyshev numerical grid.  Larger number of points are required
for smaller values of the magnetic Prandtl number.

We shall concentrate on and follow here only one mode and call it the
{\em fundamental} one. This is the mode which first becomes unstable
when the vertical magnetic field is decreased below threshold (the
mode is marginal at threshold). For given values of the parameters,
the eigenvalue corresponding to this fundamental mode arises as one of
the four possible solutions of the dispersion relation. It is purely
real ($Im(s)=0$) and thus the instability is steady, or
non-oscillatory (in the customary nomenclature, e.g., \cite{cross}).
The solution of the dispersion relation provides the functional
dependence $s = s(k,q,{\cal C},{\cal P}_m,{\cal R}_{m})$.  \par In
Figure (\ref{linear_plottery_Pm001}-a) we display the growth rate
$Re(s)$ as a function of $k$ of this fundamental mode, for several
values of ${\cal R}_m$. The parameters ${\cal C}, {\cal P}_m$ and $q$
are fixed at the values indicated in the caption.  We see that the
transition into instability is typical of steady-cellular
instabilities (similar, in principle, to Rayleigh-B\'{e}nard
convection). The marginal mode can be chosen to have a transition to
instability at the maximum of the curve $s(k)$ (i.e. $s=0$
simultaneously with $\partial s/\partial k = 0$), while all the other
modes show strong temporal decay.  The marginal mode can be identified
with respect to a critical wavenumber $k_{\rm crit} \equiv Q$ and a
critical magnetic Reynolds number ${\cal R}_m({\rm crit}) \equiv
R_m$.\\

Figure (\ref{linear_plottery_Pm001}-b), which shows the neutral curve
($s=0$) in the ${\cal R}_m - k$ plane, also demonstrates the way in
which the critical values $Q$ and $R_m$ are determined.  These
critical parameters are in general functions of the remaining
parameters of the system, i.e. $Q = Q(q,{\cal P}_m,{\cal C})$ and $R_m
= R_m(q,{\cal P}_m,{\cal C})$.  From here on out we will restrict our
considerations to values of $q=3/2$ (for consistency with UMR06) and
consider the behavior of these quantities as a function of ${\cal C}$
and, primarily, ${\cal P}_m$.  \par
\begin{figure}
\begin{center}
\leavevmode \epsfysize=7.cm
\epsfbox{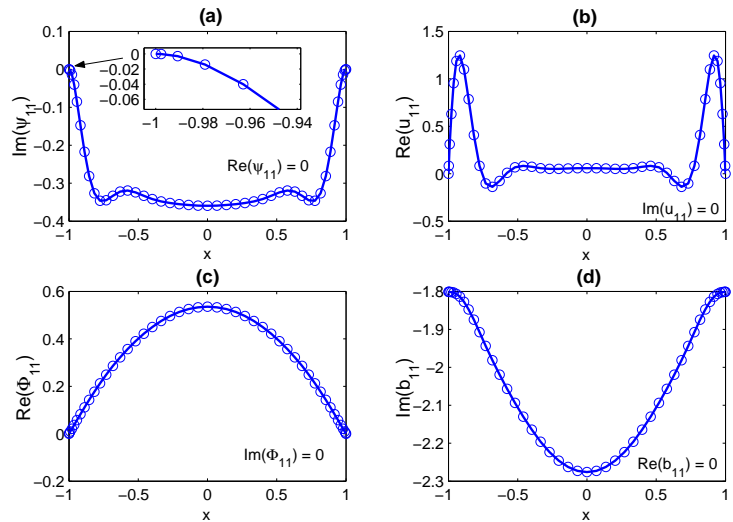}
\end{center}
\caption{{\small (Color online) Eigenfunctions for the marginal mode at ${\cal C}
= 0.08$, ${\cal P}_m = 0.001$, $q = 3/2$.  Here $Q\sim 0.75$,
$R_{_m}\sim 4.9$.  The eigenfunctions are shown fitted (solid
line) to the values determined numerically on the Chebyshev grid
(open circles).  (a)  $\Psi_{_{11}}$, (b) $u_{_{11}}$, (c)
$\Phi_{_{11}}$, (d) $b_{_{11}}$.  Note that $Re(\Psi_{_{11}}) =
Im(u_{_{11}}) = Im(\Phi_{_{11}}) = Re(b_{_{11}}) = 0$. Because the
Prandtl number is small, note rather sharp boundary layers
appearing in $\Psi_{_{11}}$ and $u_{_{11}}$. The inset in (a)
resolves the boundary layer behavior near $x=-1$. }}
\label{marginal_eigenfunctions}
\end{figure}

 The eigenfunctions for the mode in
question have even symmetry with respect to $x=0$ due to both the
symmetry in the boundary conditions and the symmetries inherent to the
thin-gap limit of the mTC problem.  In Figure
(\ref{marginal_eigenfunctions}) we display a sample of eigenfunctions
of the marginal mode.  To avoid later notational ambiguity, the
eigenfunctions for these marginal modes (i.e. those with $k=Q$ and
${\cal R}_{m} = R_m$) will be labeled with a ``11" subscript, that
is, those modes will be represented by
\[
\Psi_{_1} \leftrightarrow \Psi_{_{11}},\quad
u_{_1} \leftrightarrow u_{_{11}},\quad
\Phi_{_1} \leftrightarrow \Phi_{_{11}},\quad
b_{_1} \leftrightarrow b_{_{11}},\qquad
{\rm when} \ \  k=k_{\rm crit} \equiv Q, \ \ {\cal R}_{m} = {\cal R}_{m}({\rm crit}) \equiv  R_m.
\]
It is argued in Appendix \ref{boundary_layer_argument} that in the
limit ${\cal P}_m \ll 1 $, the boundary layers size that appear scale
as ${\cal P}_m^{1/3}$.  The boundary layers that develop are
satisfactorily represented numerically by the Chebyshev method used,
e.g. with a grid of 50 points we can resolve at least 3-4 points of
the boundary layer zones when ${\cal P}_m=10^{-5}$.  This dependence
on ${\cal P}_m$ will also have some bearing on the scaling properties
of the coefficients of the resulting (real GLE) envelope equation,
presented in the next section.  \par Finally, we note that there
always exists an additional marginal mode of the system, separate from
the above mentioned MRI mode.  This neutral mode reflects a symmetry
introduced into the system due to the conducting boundary conditions.
Namely, a spatially constant, time-independent solution for the
azimuthal magnetic field (i.e. $b_y = {\rm constant}$) solves both the
linear (and, incidentally, the nonlinear) equations and satisfies its
requisite boundary conditions.  This mode must be formally included in
the subsequent nonlinear analysis.

\section{Weakly Nonlinear Asymptotic Analysis}\label{weakly_nonlinear}
The weakly nonlinear analysis aims to develop a description of the
system's evolution beginning very close to marginality, slightly into
the unstable region. The control parameter in the asymptotic analysis
is incorporated in the expression for the background magnetic
field. It is here set to be $B_0 =1-\epsilon^2$, i.e.  the degree of
departure from marginality is controlled by the small parameter
$\epsilon$ (of our choosing) whose only \emph{formal restriction} is
that it be $\epsilon \ll 1$.

Close to marginality the relevant MRI mode, discussed in the previous
section, may be expressed to leading order in $\epsilon$ (as can be
shown by a simple scaling and balancing analysis) in the form
\[
\epsilon {\bf V}_1\ = \epsilon \left( A \mathbb{V}_{_{11}} e^{iQz} + B \mathbb{U}_{_{11}} + {\rm c.c} \right),
\]
where $\mathbb{V}_{_{11}} \equiv
(\Psi_{_{11}},u_{_{11}},\Phi_{_{11}},b_{_{11}})^\mathbf{T}$ and
$\mathbb{U}_{_{11}} \equiv (0,0,0,1)^\mathbf{T}$. The inclusion of $B
\mathbb{U}_{_{11}}$ in this general solution is dictated by the
presence of the neutral mode, discussed at the end of the previous
Section.

The weakly nonlinear evolution is asymptotically derived by
allowing the amplitudes $A$ and $B$ to be (weakly) dependent on
space and time. The aim is to develop an evolution equation for
the envelopes $A$ and $B$ (space and time dependent amplitudes) as
one tunes the system away from the marginal state defined above at
$k=Q$ and ${\cal R}_{m} = R_m$.  The wisdom garnered from other
problems involving cellular instabilities
\cite{manneville1,cross,manneville2,pismen} guides us into an
Ansatz such that the two envelope functions have functional
dependencies upon a long time scale, $T \equiv \epsilon^2 t$ and a
long vertical scale, $Z \equiv \epsilon z$, i.e. we posit the form
$A=A(\epsilon^2 t,\epsilon z), B=B(\epsilon^2 t,\epsilon z)$. The
end-result of this asymptotic procedure, fully detailed in
Appendices B and C, are the two (decoupled) amplitude equations
\beqa
\partial_{_T} A &=& \lambda A - \alpha A|A|^2 + D \partial^2_{_Z} A ,
\label{landau_eqn}\\ \partial_{_T} B &=&
\left(\frac{1}{R_{_m}}+\frac{{\cal C}{\cal R}}{3}\right)
\partial^2_{_Z} B,\label{diffusion_eqn}
\eeqa
 where $T \equiv \epsilon^2 t$, $Z \equiv \epsilon z$ and the
coefficients are defined in Appendix B.

We stress here that the decoupling of these two equations is the
result of translational ($x$-) symmetry of the thin-gap problem, but
it cannot be guaranteed for a case in which, e.g., curvature terms
have to be retained.  Eq. ({\ref{diffusion_eqn}) is the diffusion
equation and its physical implications are quite trivial.  It
indicates that the contribution of the above mentioned neutral mode to
the azimuthal field perturbation will simply decay on a time-scale
associated with the system's size and the smaller of either $R_m$ or $1/{\cal C}{\cal R}$
- the meaning of the latter possibility will be explored in a forthcoming work.
In contrast,
equation (\ref{landau_eqn}) is the well-studied real Ginzburg-Landau
equation (see, e.g. \cite{cross,manneville2,pismen})
which can exhibit non-trivial behavior in both the amplitude and phase
of the envelope function $A$.  The phase can lead to interesting
dynamics emerging from Eckhaus-like instabilities, however in the
present study we care only about the behavior of the amplitude's
magnitude, i.e. the modulus of A.  We shall thus agree henceforth to
mean $|A|$, when writing $A$.  Further discussion on phase dynamics
can be found in the concluding section of this paper.  \par A real
amplitude $A$ in the real GLE has two stable spatially uniform steady
solutions, $A(Z,T)=\pm A_s$, and one possibly unstable solution,
$A=0$, as can be easily verified.  Depending on the boundary
conditions, the system typically relaxes to one of the steady
solutions or, possibly, splits into two regions (the plus and minus
values of $A_s$) with a front separating them (see e.g., \cite{regev}
for an example of a system of this kind).
\begin{figure}
\begin{center}
\leavevmode \epsfysize=6.5cm
\epsfbox{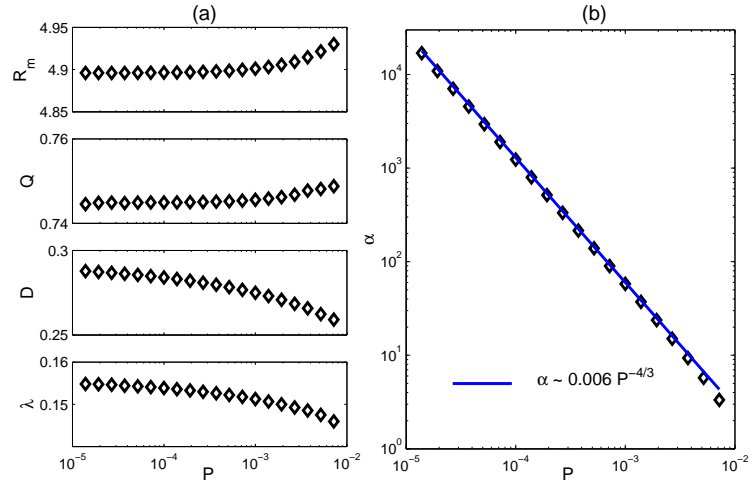}
\end{center}
\caption{ {\small (Color online) Coefficients, parameters and saturation amplitude
squared, as a function of ${\cal P}_m$, for $q=3/2$, ${\cal C} =
0.08$.  (a) Plots of $R_m$, $Q$, $D$, $\lambda$ as a function of
${\cal P}_m$.  Note the weak sensitivity on ${\cal P}_m$. (b) Values
of $\alpha$ (diamonds) as a function of ${\cal P}_m$.  Plotted as well
is the functional dependence $\alpha \sim {\cal P}_m^{-4/3}$ (solid
line) discussed in the text.}}
\label{coefficient_plots}
\end{figure}

From (\ref{landau_eqn}) it is apparent that the saturation amplitude
is $A_s = \sqrt{\lambda/\alpha}$ and thus its determination calls for
the computation of the relevant coefficients. As discussed before this
has to be done numerically.  The details of this calculation are given
in Appendix B and some representative results
(for the parameter values $q=3/2$ and ${\cal C}=0.08$) are displayed
in Figure \ref{coefficient_plots}. Panel (a) demonstrates the weak
dependence of $R_m$ and $Q$, and of the coefficients $\lambda$ and
$D$, on ${\cal P}_m$ (for ${\cal P}_m\ll 1$). In contrast, the
coefficient $\alpha$, whose numerical values are shown in panel (b),
has a power law dependence on ${\cal P}_m$ in the same interval. Thus,
the dependence of the saturation amplitude on ${\cal P}_m$ is
essentially governed by $\alpha$. The appropriate scaling for ${\cal
P}_m \ll 1$ is $A_s^2=\lambda/\alpha \sim 1/\alpha$ (given the very
weak dependence of $\lambda$ on ${\cal P}_m$).

\begin{figure}
\begin{center}
\leavevmode \epsfysize=10.5cm
\epsfbox{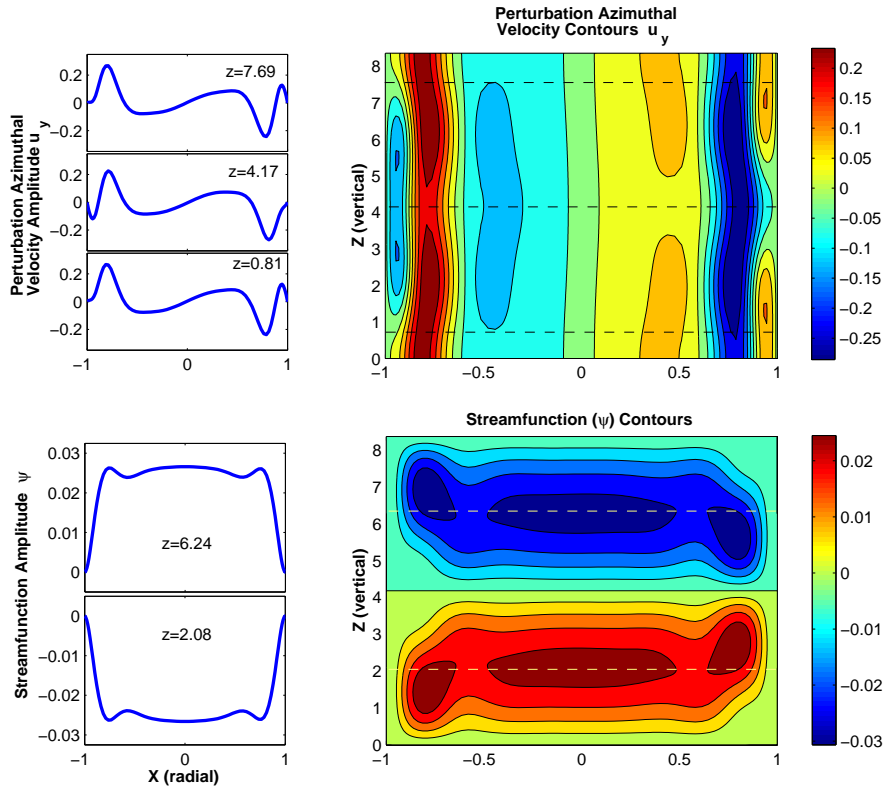}
\end{center}
\caption{{\small (Color online) Contours of the perturbation streamfunction $(\Psi)$
and azimuthal velocity $(u_y)$ in the $x-z$ plane, up to and including
order $\epsilon^2$.  The vertical scale of the plots corresponds to
one critical wavelength, $2\pi/Q$.  The parameters are ${\cal P}_m =
0.001$, ${\cal C} = 0.08$, $\rm R_m = 4.9$ with $q=3/2$.  We take
$\epsilon = 0.5$ and the amplitude $A = 0.07$ (which is its saturation
value for this case).  Cuts along constant values of $z$ are shown in
the left panel (and dashed lines in the contour plots on the right ).
Note that $u_y$ is the velocity disturbance about the steady profile
$-qx$.\\ }} \label{boundary_layer_fig1}
\end{figure}

The analysis sketched out in Appendix D shows
that the dominant terms in the expression for $\alpha$ are such
that $\alpha \sim  {\cal P}_m^{-4/3}$, for ${\cal P}_m \ll 1$
This scaling fits very well the numerical results in
Figure \ref{coefficient_plots}b (solid line). We thus obtain the
following scaling behavior for the square of the saturation amplitude
\beq
A_s^2 \sim  {\cal P}_m^{4/3}~({\rm or}~~ A_s^2 \sim {\cal R}^{-4/3}~{\rm for~fixed}~
R_m),
\qquad {\rm both~for} \ \ {\cal
P}_m \ll 1. \label{alpha_scaling_behavior}
\eeq

The physical effects that these dominant terms are reflecting can be
traced in the asymptotic analysis as resulting from the nonlinear
radial advection of the second order azimuthal velocities $u_{x1}
\partial_x u_{y2}$ and the creation of the azimuthal field due to the
shearing of the radial perturbation field $b_{x1} \partial_x
u_{y2}$. Note that in UMR06 we were able to obtain (from not fully
consistent boundary conditions for this problem) the analytical result
$A_s^2 \sim {\cal P}_m$ (or $\sim {\cal R}^{-1}$ for fixed $R_m$).
Thus we see that the implementation of more realistic boundary
conditions that are appropriate for the thin-gap mTC problem does not
alter the general {\em qualitative} trend - saturation amplitude
increasing with ${\cal P}_m$ (or decreasing with $\cal R$ for fixed
$R_m$) - uncovered in UMR06, nor its implications. It merely alters
(slightly) the power of this basic dependence.

In Figure \ref{boundary_layer_fig1} we plot the azimuthal velocity
$u_y(x,z)$ and the streamfunction $\Psi(x,z)$ of the perturbation,
calculated by our asymptotics to order $\epsilon^2$. This has to be
understood as the modification on top of the basic mTC configuration,
which together constitute the steady saturated state.  The presence of
boundary layers near the channel walls is clearly apparent.  In
Appendix \ref{boundary_layer_argument} we estimate that the boundary
layer sizes scale as $\sim {\cal P}_m^{1/3}$ and this is
quantitatively consistent with the increase in power of the scaling
from $A_s^2 \sim {\cal P}_m$ (as found in UMR06, where the boundary
layers were essentially neglected) to $A_s^2 \sim {\cal P}_m^{4/3}$
here. The crucial ingredient in determining the scaling of $A_s$ is,
as we have seen, the scaling behavior of the coefficient $\alpha$,
which in turn is affected by the boundary layer width through its
dependence on the relevant $x$-eigenfunctions (see Appendices A and D).

\begin{figure}
\begin{center}
\leavevmode \epsfysize=10.5cm
\epsfbox{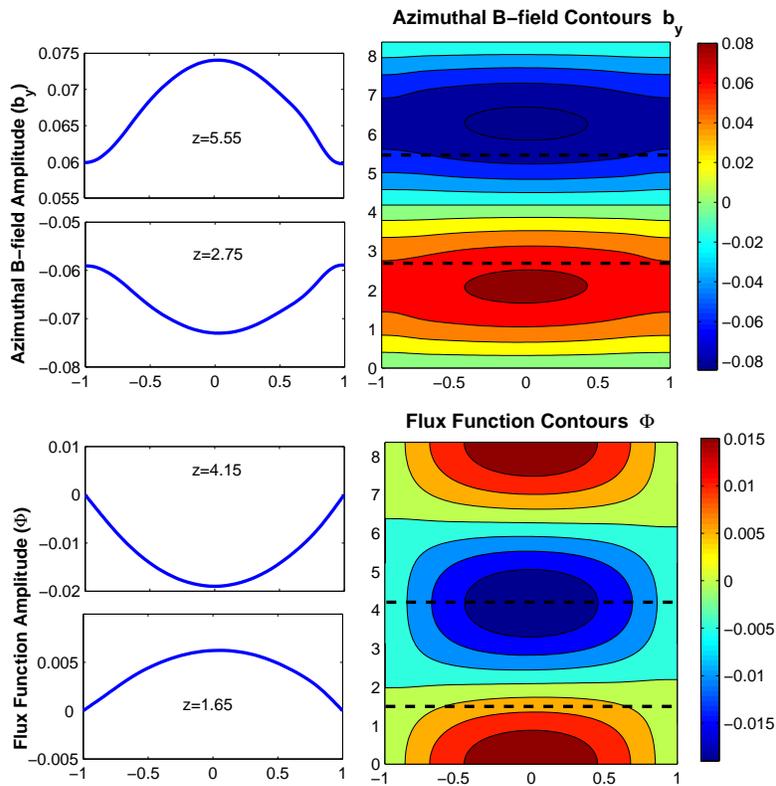}
\end{center}
\caption{{\small (Color online) Same as Fig. \ref{boundary_layer_fig1} for the
azimuthal field (top panels), $b_y$, and the flux function, $\Phi$
(bottom panels).
}} \label{B_fig}
\end{figure}
\par

In Figure \ref{B_fig} we display the perturbation's azimuthal field,
$b_y$, and its poloidal flux function, $\Phi(x,z)$, in a manner
similar to the previous figure. Note that we do not see prominent
boundary layers in the magnetic field perturbation; this is the result
of the boundary conditions imposed (\ref{bcc}). Whereas three velocity
boundary conditions are imposed on each side (ensuring zero
perturbation velocity on the boundary), only two such conditions on
the magnetic field perturbation are enforced ($b_x=0, \partial_x b_y
=0$). It is so because precisely ten conditions in all are required,
otherwise the problem would be ill-posed.

\begin{figure}
\begin{center}
\leavevmode \epsfysize=9.5cm
\epsfbox{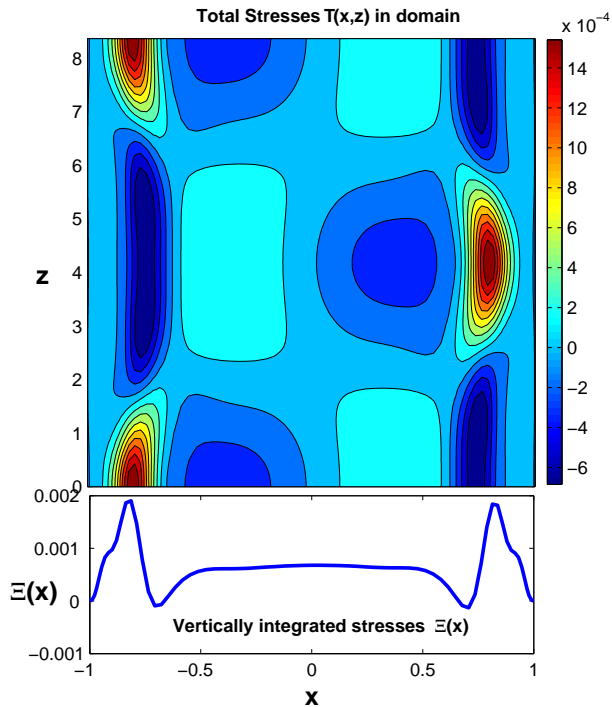}
\end{center}
\caption{{\small (Color online) Same as Fig. \ref{boundary_layer_fig1} for the total
stress, ${\mathbb T}(x,z)$ (top panel), and the vertically integrated
stress, $\Xi=\int{\mathbb T}(x,z)dz$ (bottom panel).
}} \label{T_fig}
\end{figure}
\par

Finally, we turn to the evaluation of the angular momentum transport
(a key question in assessing the MRI's role as the driver of accretion
in astrophysical systems). The total (local) stress resulting from the
perturbation is composed of the Reynolds and Maxwell stresses and, in
our notation, has the form (see, e.g., \cite{B03},\cite{pesahMN},
UMR06) \beq \mathbb{T}(x,z) \equiv \mathbb{T}_{\rm R}+\mathbb{T}_{\rm
M}=u_x u_y - {\cal C} b_x b_y.
\label{RMstresses}
\eeq

As in UMR06, we may define a quantity measuring the average total
angular momentum transport in the domain,
\[ {\mathbf {\dot J}} = \frac{Q}{4\pi}\int_{-\pi/Q}^{\pi/Q}dz \left[ \int_{-1}^1
\mathbb{T}(x,z) dx \right].
\]
The quantity ${\mathbf {\dot J}}$ can be thought of as analogous to
similarly defined quantities used as a measure of the effective
viscosity parameter due to dynamical fluctuations in active fluid
media (for a recent purely hydrodynamic example, see
\cite{longaretti2}) - be it either in a fully turbulent state or
otherwise. In our problem, this quantity in the saturated state can be
written to leading order as \beq {\mathbf {\dot J}} = \epsilon^2 |A|^2
\dot J_{0} + \order{\epsilon^4}, \eeq where $\dot J_{0} = \dot J_{0}(
R_m,Q; {\cal C},q)$ is of order unity.  Given the behavior of the
saturated amplitude in the ${\cal P}_m \ll 1$ limit, it follows that
the average angular momentum transport scales like \beq {\mathbf {\dot
J}} \sim \epsilon^2{\cal P}_m^{4/3}~~ ({\rm or}~ \sim \epsilon^2{\cal
R}^{-4/3}~ {\rm for~fixed}~ R_m~{\rm of}~ \order{1} ) , \eeq to
leading order.  Finally, we show in Figure \ref{B_fig} the distributed
stress $\mathbb{T}(x,y)$ over the domain and the vertically integrated
stress, defined by $ \Xi(x) \equiv \int \mathbb{T}dz$.

\section{Discussion and summary}\label{Discussion}

In this paper we have presented a full exposition of a weakly
nonlinear asymptotic analysis of the MRI for a viscous and resistive
flow in the thin-gap magnetic Taylor-Couette configuration. Our
previous work (UMR06) employed mathematically expedient, but not fully
consistent boundary conditions for this problem, so as to allow for
transparent analytical evaluation of the envelope equation
coefficients.  Here we have used consistent and realistic boundary
conditions for the mTC setup. As a result, the calculation is more
involved. We have nevertheless found (as anticipated in UMR06) that in
the thin gap limit the amplitude of the disturbances saturates at a
value that decreases with decreasing magnetic Prandtl number, ${\cal
P}_m$. Moreover, the emergence of boundary layers actually makes the
${\cal P}_m$-dependence of the saturation amplitude, and thus the
average angular momentum transport more severe.  \par Our results
should be put in the proper context. They are valid {\em close to
instability threshold} and in a {\em confined system} (mTC). Most
previous studies of the MRI in the nonlinear regime (both numerical
and analytical) followed the evolution of {\em channel modes} -
exponentially growing, radially independent modes (see \cite{BHrev}),
which happen also to be exact solutions of the nonlinear equations for
the perturbation, in the SB formulation under {\em periodic} boundary
conditions, i.e.  in an {\em open} system. It is thus only natural
that the channel modes have been identified as the dominant dynamics
and their evolution perceived as a crucial ingredient in the nonlinear
saturation of the instability. Goodman \& Xu \cite{GX} showed that the
channel modes ultimately become unstable and break up. The asymptotic
study of nonlinear saturation performed by Knobloch \& Julien
\cite{knobloch} was also based on a state dominated by channel
modes. In these works, as well as the recent local modeling of MRI
angular momentum transport \cite{pesahMN,pesahPRL}, results were
compared with numerical simulations of an open SB (undoubtedly
dominated by dynamics arising from the nonlinear evolution of channel
modes). Note, however, that in the global approach of Kersal\'e et
al. \cite{kersale1,kersale2}, the explicit inclusion of boundary
conditions and curvature terms broke the radial symmetry of the
problem (which is necessary for the channel modes to be manifested).
These authors found numerically (using a spectral code) that the form of the saturated state
critically depends on the boundary conditions adopted and, in any
case, is not a ``trivial" Keplerian state with developed MHD turbulence
on top of it.

From the vantage point of the linear theory followed here (as well as
the SB investigations of the past), the MRI takes place primarily
because the term supplying the tension, i.e. a perturbed azimuthal
$B$-field, arises from the sheared conversion (by the background flow)
of a perturbed radial magnetic field, emanating from the bending of the
background vertical field. The strength of the resulting destabilizing
torque is related to the magnitude of $q$ (measuring the local
stretching) and the magnitude (squared) of the global vertical
$B$-field (representing that basic source of tension which is being
stretched by the shear). Nonlinear saturation of a linear instability
can generically be achieved by increased dissipation, by the
modification of the linearly unstable base state so as to push it back
to stability, or a combination of both.

In the problem studied here, we have considered the marginal MRI mode
(i.e. with growth rate 0), as a function of all free parameters, {\em
save} $q$, which has been fixed to 3/2. We find that the saturated
azimuthal velocity disturbance provides an effective {\em positive}
radial gradient, $q'>0$, through the bulk of the flow (see Figure
\ref{boundary_layer_fig1}). Thus the effective overall $q$ in the
saturated state is $q_{\rm eff}=q-q'<3/2$.  The magnitude of the
effective gradient reflects the manner in which the modified gradient
couples to the background field which is being stretched and is
responsible for the instability. In our case, $q'$ is positive and
thus reduces the initial destabilizing shear, but not sufficiently to
cancel it entirely. It has to be noted, however, that the saturated
state is {\em not} just the base flow with reduced shear.  It includes
also extra poloidal and azimuthal field, as well as poloidal
velocity. This steady state is thus more complicated; the presence of
velocity boundary layers complicates it even further.  It is thus not
trivial to identify a simple process for the saturation "mechanism" in
this case.  We note that our results share similarities with the
saturation mechanism proposed by Knobloch \& Julien \cite{knobloch}
for the saturated MRI state developed, in a particular asymptotic
regime, from the unstable channel modes discussed above.\par

We have followed into the weakly nonlinear regime a dissipative
system, which was in a marginal balance and obtained a steady
saturated state from a reduction of the shear, in places over the
domain where it counts the most (in terms of azimuthal field
production), and from the emergence of a steady flow and magnetic
field configuration.  In terms of dissipation, it is instructive to
consider the energy relationship (\ref{energy_theorem}).  In our
steady saturated state the first integral is just $\propto
\int\Xi(x)dx$ and therefore is positive (see the bottom of
Fig. \ref{T_fig}). As $\partial_t E$ in this (steady) saturated state
must be zero, the sum of the two dissipative integrals must be equal
to the first one. We have verified that it is indeed so.

We have not considered in this paper phase dynamics, which is an
inherent feature of the more general envelope in complex GLE.  Phase
dynamics may be rich, in particular in two and three dimensions,
admitting well-known pattern instabilities like Eckhaus and Zig-Zag
and these, in turn, can lead to effects like phase turbulence and
complicated defect dynamics \cite{manneville2,pismen}.  In what is
considered here, where the coefficients of the one-dimensional GLE are
real, all that remains of the above is just a possibility of an
Eckhaus instability. This may merely introduce some non-steady
readjustment to the overall pattern phase, but it leaves unaltered the
overall amplitude scale of the basic pattern that emerges. In
particular, our system is open in the $z$ dimension and thus there
should be no difficulty for the phase to adjust itself to a stable
value (see \cite{manneville2} p. 200).  Because we are interested here
in the scaling of the transport (which is expressed by an integral of
the envelope over the domain), phase dynamics (although interesting in
its own right) does not influence this measure and we have thus
considered only the modulus of the envelope.  \par

Our results and findings here should ultimately be compared to
experiments and numerical simulations accompanying them. Extension
of this type of analysis to a wide-gap mTC configuration is
possible, but the results will be somewhat more complicated than
those presented here, due to the inclusion of curvature terms.
Preliminary calculations indicate that the evolution of the
perturbation amplitude in this case is governed by two coupled
envelope equations (see Appendix B). The properties of the
saturated state, however, appear similar in their salient features
to the ones explored in this paper. The case of an initial helical
field, for which experimental detection of the MRI has recently
been reported \cite{helical}, can also be investigated in the
weakly nonlinear asymptotic formalism employed here. It will the
subject of future work.

Further analytical investigations of the nonlinear MRI, of the kind
reported here, will contribute toward assembling a deeper
understanding of this important instability.  Such investigations may
also help in addressing the issues of the effect of numerical
resolution upon the resulting dynamics. In particular it could be
useful to conduct simulations for, say, a fixed value of the magnetic
Reynolds number (well below any contamination by numerical
dissipation) and examine if and how does the transport change with
resolution.  Numerical studies of the MHD turbulent dynamo problem
(e.g., \cite{cata,brand2}) have shown that such considerations are
very important.  The understanding of the role that the MRI plays in
astrophysical disks, which in its full generality is a formidable
problem, may be enriched by the experimental, analytical and numerical
studies of simpler systems.
\vskip.5cm
\acknowledgements
The authors would like to thank the Israel Science Foundation
and BSF grant number 0603414082 for partial support of this study.  We are
greatly indebted to G. Shaviv and E. A. Spiegel for sharing with
us their comments
and insights.  In addition we thank the two anonymous referees whose
comments helped to improve the presentation of this work.

\appendix

\section{The linear scale of the boundary layer}
\label{boundary_layer_argument}
The best way to identify the scalings that are appropriate for the
boundary layer is to rewrite (\ref{linear_operator}) as a single
equation for, say, the streamfunction $\Psi$.  Setting the
time-derivative to zero results in \beq {\cal L}\Psi = \left\{
\left({\cal P}_m(D_x^2-Q^2)^2 + {\cal C} R_m^2 Q^2\right)^2(D_x^2 -
Q^2) +R_m^4 2q{\cal C}Q^4 - \omega_e^2 Q^2(D_x^2-Q^2)^2 {R_m^2}
\right\}\Psi = 0,
\label{single_operator_L}
\eeq
where $\omega_e^2 \equiv 2(2-q)$ and where the simplifying notation $D_x\equiv d/dx$ is
also used.
The operator is tenth order in $D_x$ derivatives.  Inspection of its
form suggests that retaining only the terms of
(\ref{single_operator_L}) that are dominant (for ${\cal P}_m\ll 1$) in
a small region of size $~{\cal P}_m^{\lambda}$ with $\lambda>0$ (the
total $x$-domain size is 2 in our units) at either of the two
boundaries gives \beq \left({\cal P}_m^2 D_x^{10} - \omega_e^2
Q^2D_x^4 {R_m^2}\right)\Psi = 0.  \eeq Treating all quantities as
being of $\order{1}$ except for ${\cal P}_m$, we can now see that the
value of the exponent $\lambda$ must be $1/3$.  More explicitly, we
consider a boundary layer by rescaling the x-coordinate around the
boundaries at $x=\pm 1$.  We define $\xi \equiv {\cal
P}_m^{-\lambda}(x\mp 1)$ and insert this into
(\ref{single_operator_L}) revealing \beq \left({\cal
P}_m^{2-10\lambda} D_\xi^{10} - {\cal P}_m^{-4\lambda} \omega_e^2
Q^2{R_m^2}D_\xi^4\right)\Psi + \order{{\cal
P}_m^{2-8\lambda},\cdots,1} = 0, \eeq where $D_\xi \equiv d/d\xi$.  As
${\cal P}_m \rightarrow 0$, a distinguished balancing limit (see,
e.g. \cite{bender}) may be achieved when $2-10\lambda = -4\lambda$, or
when $\lambda = 1/3$.  In this case, all other terms in the boundary
layer region are sub-dominant to the two terms remaining.  Thus, in the
limit ${\cal P}_m \ll 1$, the size of the boundary layer scales as
${\cal P}_m^{1/3}$.\\

\section{Derivation of the Ginzburg-Landau Equation}\label{landau_derivation}

To help facilitate the development of the weakly nonlinear theory
we rewrite the equations of motion in the following way, \beq {\bf
{\cal D}} \partial_t {\bf V} + {\bf N} = {\bf {\cal L}}{\bf V} +
\epsilon^2 {\bf {\cal G}} {\bf{V}}, \label{fundamental_equation}
\eeq in which \beq {\bf {\cal L}} \equiv {\bf { L}_0} +{\bf {
L}_{1}}\partial_z +{\bf { L}_{2}}\partial_z^2 +{\bf {
L}_{3}}\partial_z^3 +{\bf { L}_{4}}\partial_z^4,
\label{L_Definition} \eeq and where the matrices are defined as
\beqa {\bf { L}_0} &=& \left(
\begin{array}{cccc}
{\cal R}^{-1}\partial_x^4 & 0 & 0 & 0 \\
0& {\cal R}^{-1}\partial_x^2  & 0 & 0 \\
0&0&{\cal R}_{m}^{-1}\partial_x^2  & 0 \\
0&0&0&{\cal R}_{m}^{-1}\partial_x^2
\end{array}
\right),\qquad {\bf {\cal D}} = \left(
\begin{array}{cccc}
\partial_x^2+\partial_z^2 & 0 & 0 & 0 \\
0& 1  & 0 & 0 \\
0&0&1  & 0 \\
0&0&0&1
\end{array}
\right), \nonumber \\
 {\bf { L}_{1}} &=&
\left(
\begin{array}{cccc}
0 & 2 & {\cal C}\partial_x^2 & 0 \\
2-q& 0 & 0 & {\cal C} \\
1&0&0  & 0 \\
0&1&-q&0
\end{array}
\right), \qquad {\bf { L}_{2}} = \left(
\begin{array}{cccc}
2{\cal R}^{-1}\partial_x^2 & 0 & 0 & 0 \\
0& {\cal R}^{-1}  & 0 & 0 \\
0&0&{\cal R}_{m}^{-1}  & 0 \\
0&0&0&{\cal R}_{m}^{-1}
\end{array}
\right), \nonumber \\
{\bf { L}_{3}} &=& \left(
\begin{array}{cccc}
0 & 0 & {\cal C} & 0 \\
0& 0  & 0 & 0 \\
0&0&0  & 0 \\
0&0&0&0
\end{array}
\right),\qquad {\bf { L}_{4}} = \left(
\begin{array}{cccc}
{\cal R}^{-1} & 0 & 0 & 0 \\
0& 0  & 0 & 0 \\
0&0&0  & 0 \\
0&0&0&0
\end{array}
\right), \qquad {\bf {\cal G}} = \left(
\begin{array}{cccc}
0 & 0 & {\cal C}\partial_x^2 & 0 \\
0& 0 & 0 & {\cal C} \\
1&0&0  & 0 \\
0&1&0&0
\end{array}
\right). \label{matrix_definitions} \eeqa At marginality, the
background vertical field $B_0=1$.  The degree of linear
instability is thus governed by the small parameter $\epsilon$
defined by \beq
 \epsilon^2 \equiv 1-B_0.
\eeq The vectors above are defined by
\[
{\bf V} \equiv (\Psi,u,\Phi,b)^\mathbf{{T}},\qquad {\bf N} \equiv
(N_\Psi,N_u,N_\Phi,N_b)^\mathbf{T}.
\]
We assume that during the nonlinear development there are two
vertical scales emerging in the problem, namely $z$ and $Z \equiv
\epsilon z$. We also assume that as one tunes the vertical field
parameter $\epsilon^2$ into the MRI unstable state, the temporal
response scales similarly to $\epsilon^2$.  Thus we say, for
example for the streamfunction, that $\Psi = \Psi(z,\epsilon z,
\epsilon^2 t)$, and similarly for the other physical variables.
If we assume that the solution forms follows ${\bf V} = {\bf
V}(x,z,Z,T)$, then all operators in (\ref{fundamental_equation})
are re-expressed by applying the replacements \beq \partial_z
\longrightarrow \partial_z + \epsilon \partial_{_Z}, \qquad
\partial_t \longrightarrow \epsilon^2
\partial_{_T}.  \eeq Thus, we now have \beq \epsilon^2{\mathbf{{\cal
D}}}\,\partial_{_T}{\mathbf{V}} +{\mathbf{N}} = {\cal
L}{\mathbf{V}} +\epsilon\mathbf{\tilde{{\cal
L}}_1}\partial_{_Z}{\mathbf{V}} +\epsilon^2\mathbf{\tilde{{\cal
L}}_2}\partial^2_{_Z}{\mathbf{V}} +\epsilon^2\mathbf{\tilde{{\cal
G}}}{\mathbf{V}} + \order{\epsilon^3}, \eeq where \beqa
\mathbf{\tilde{{\cal L}}_1} &=& {\bf { L}_{1}}+ 2{\bf {
L}_{2}}\partial_z +3{\bf {L}_{3}}\partial_z^2 +4{\bf {
L}_{4}}\partial_z^3, \\ \mathbf{\tilde{{ \cal L}}_2} &=& 2{\bf {
L}_{2}} +6{\bf { L}_{3}}\partial_z +12{\bf { L}_{4}}\partial_z^2.
\eeqa The boundary conditions, aside from periodicity in the
vertical, are \beq \partial_z\Psi = \partial_x\Psi = u_y =
\partial_z\Phi =
\partial_x b_y = 0, \qquad {\rm at} \ \ x = \pm 1.  \eeq We expand all
quantities in a perturbation series \beqa \Phi &=& \epsilon\Phi_1
+ \epsilon^2\Phi_2 + \epsilon^2\Phi_3 + \cdots \nonumber \\ \Psi
&=& \epsilon\Psi_1 + \epsilon^2\Psi_2 + \epsilon^2\Psi_3 + \cdots
\nonumber \\ u_y &=& \epsilon u_1 + \epsilon^2 u_2 + \epsilon^3
u_3 + \cdots \nonumber \\ b_y &=& \epsilon b_1 + \epsilon^2 b_2 +
\epsilon^3 b_3 + \cdots \nonumber \eeqa or in other words
\[
{\mathbf{V}} = \epsilon{\mathbf{V}}_1 + \epsilon^2{\mathbf{V}}_2+
\epsilon^3{\mathbf{V}}_3 + \order{\epsilon^4}.
\]
With the above expansion and multiple scaling Ansatz, it follows
that the nonlinear terms are expressed in the series \beq
{\mathbf{N}} = \epsilon^2 {\mathbf{N}_{2}} +\epsilon^3
{\mathbf{N}_{3}} + \order{\epsilon^4}, \eeq in which
${\mathbf{N}}=(N^{(\Psi)},N^{(u)},N^{(\Phi)},N^{(b)})^{\mathbf{T}}$.
In component by component form, these expressions are explicitly
given by \beqa N^{(\Psi)} &=& \epsilon^2 N^{(\Psi)}_{2} +
\epsilon^3 N^{(\Psi)}_{3} +
\order{\epsilon^4} \\
N^{(u)} &=& \epsilon^2 N^{(u)}_{2} + \epsilon^3 N^{(u)}_{3} +
\order{\epsilon^4} \\
N^{(\Phi)} &=& \epsilon^2 N^{(\Phi)}_{2} + \epsilon^3
N^{(\Phi)}_{3} +
\order{\epsilon^4} \\
N^{(b)} &=& \epsilon^2 N^{(b)}_{2} + \epsilon^3 N^{(b)}_{3} +
\order{\epsilon^4} \eeqa where \beqa N^{(\Psi)}_{2} &=&
J(\Psi_1,\nabla^2\Psi_1) - {\cal C}J(\Phi_1,\nabla^2\Phi_1)
\label{F_psi_1_def}
\\
N^{(u)}_{2}
&=& J(\Psi_1,u_1) - {\cal C}J(\Phi_1,b_1), \\
N^{(\Phi)}_{2}
&=& - J(\Phi_1,\Psi_1), \\
N^{(b)}_{2} &=& J(\Psi_1,b_1) - J(\Phi_1,u_1), \label{F_u_1_def}
\eeqa and \beqa N^{(\Psi)}_{3} & = &  J(\Psi_2,\nabla^2\Psi_1) +
J(\Psi_1,\nabla^2\Psi_2)- {\cal C}J(\Phi_2,\nabla^2\Phi_1)
-{\cal C}J(\Phi_1,\nabla^2\Phi_2) \nonumber \\
& & \tilde J(\Psi_1,\nabla^2\Psi_1) -{\cal C}\tilde
J(\Phi_1,\nabla^2\Phi_1)
 -2
J(\Psi_1,\partial_z\partial_{_Z}\Psi_1)
-2J(\Phi_1,\partial_z\partial_{_Z}\Phi_1),
\\
N^{(u)}_{3} &=& J(\Psi_2,u_1) + J(\Psi_1,u_2) - {\cal
C}J(\Phi_2,b_1) - {\cal C}J(\Phi_1,b_2) -
 +\tilde
J(\Psi_1,u_1) - {\cal C}\tilde J(\Phi_1,b_1),
 \\
N^{(\Phi)}_{3} &=& - J(\Phi_2,\Psi_1)- J(\Phi_1,\Psi_2) - \tilde
J(\Phi_1,\Psi_1),
 \\
N^{(b)}_{3} &=& J(\Psi_2,b_1) + J(\Psi_1,b_2) - J(\Phi_2,u_1) -
J(\Phi_1,u_2) + \tilde J(\Psi_1,b_1) - \tilde J(\Phi_1,u_1), \eeqa
in which $\tilde J(f,g) \equiv \partial_{_Z} f\partial_x g -
\partial_x f\partial_{_Z} g $ (remembering also that $ J(f,g)
\equiv \partial_z f\partial_x g - \partial_x f\partial_z g$).
\par
To $\order{\epsilon}$ we have \beq {\cal L}{\mathbf{V}_{_1}} = 0.
\eeq The solution to this is the marginal case investigated in the
text. We write its general solution form as \beq {\mathbf{V}_{_1}}
= A(T,Z)\mathbb{V}_{_{11}}e^{iQz} + {{\rm c.c.}}
 + B(T,Z)\mathbb{U}_{_{11}},
 \label{order_1_general_solution}
\eeq where $\mathbb{V}_{_{11}}(x) =
(\Psi_{_{11}},u_{_{11}},\Phi_{_{11}},b_{_{11}})^{\mathbf{T}}$,
$\mathbb{U}_{_{10}}\equiv(0,0,0,1)^\mathbf{T}$ and $A$ and $B$
(not to be confused with the magnetic field) are envelopes
(amplitudes). Keeping in mind the above cited solution form, the
boundary conditions at this order are \beq \Psi_{_{11}} = D_x
\Psi_{_{11}} = u_{_{11}} = \Phi_{_{11}} = D_x b_{_{11}} = 0,
\qquad {\rm at} \  x = \pm 1, \eeq where $D_x \equiv d/dx$. Note
that the constant azimuthal field symmetry discussed in the main
text is embodied in the final term of
(\ref{order_1_general_solution}), i.e. $B(T,Z)\mathbb{U}_{_{11}}$.
We also call attention to the fact that though all of these
functions are order 1, they all (especially $u_{_{11}}$) show the
presence of boundary layers in a region $\order{{\cal P}_m^{1/3}}$
close to the two boundaries.
\par
At order $\epsilon^2$, the equations are \beq
 {\cal L}{\mathbf{V}_{_2}} =
{\mathbf{N}_{2}} -\mathbf{\tilde{{\cal
L}}_1}\partial_{_Z}{\mathbf{V}_{_1}}. \eeq The solution at this
order is written as \beq {\mathbf{V}_{_2}} =
A^2\mathbb{V}_{_{22}}e^{i2Qz} +
\partial_{_Z} A \mathbb{V}_{_{21}}e^{iQz}
+|A|^2\mathbb{V}_{_{20}} + \partial_{_Z} B \mathbb{U}_{_{20}} +
{{\rm c.c.}}
 \label{order_2_general_solution}
\eeq where $\mathbf{V}_{_{2}} \equiv
(\Psi_{2},u_{2},\Phi_{2},b_{2})^{\mathbf{T}}$ and in particular,
$\mathbb{V}_{_{22}}(x) \equiv
(\Psi_{22},u_{22},\Phi_{22},b_{22})^{\mathbf{T}}$,
$\mathbb{V}_{_{21}}(x) \equiv
(\Psi_{21},u_{21},\Phi_{21},b_{21})^{\mathbf{T}}$,
$\mathbb{V}_{_{20}}(x) \equiv
(\Psi_{20},u_{20},\Phi_{20},b_{20})^{\mathbf{T}}$,and
$\mathbb{U}_{_{20}}(x) \equiv (0,\tilde u_{20},0,0)^{\mathbf{T}}$.
${\mathbf{N}_{2}}$ contains no terms resonant with $e^{iQz}$ (see
below) but because the expression $\mathbf{\tilde{{\cal
L}}_1}\partial_{_Z}{\mathbf{V}_{_1}}$ does contain such a term, in
order for there to be a bounded solution at this order with the
required boundary condition, the following solvability condition
(the vanishing of an {\em inner product} \cite{note_2}) must be
satisfied: \beq \left<{{\mathbf{V}^{\dagger}\cdot
\mathbf{\tilde{{\cal L}}_1}\partial_{_Z}{\mathbf{V}_{_1}}}}\right>
\equiv
\frac{Q}{2\pi}\int_{_{-1}}^{1}\int_{-\pi/Q}^{\pi/Q}dxdz{{\mathbf{V}^{\dagger}\cdot
\mathbf{\tilde{{\cal L}}_1}\partial_{_Z}{\mathbf{V}_{_1}}}} = 0,
\label{order_epsilon2_solvability} \eeq where
${\mathbf{V}^{\dagger}}$ is the solution to the adjoint operation
\beq \mathbf{{\cal L}^{\dagger}}{\mathbf{V}^{\dagger}} = 0. \eeq
The adjoint operator is given by \beq {\mathbf {\cal L}^{\dagger}}
\equiv {\mathbf { L}_0} -{\mathbf { L}_{1}}^{\mathbf{T}}\partial_z
+{\mathbf { L}_{2}}\partial_z^2 -{\mathbf {
L}_{3}}^{\mathbf{T}}\partial_z^3 +{\mathbf { L}_{4}}\partial_z^4.
\eeq and is so written since
\[
{\mathbf { L}_{0}}=\left({\mathbf { L}_{0}}\right)^{\mathbf{T}},
\quad {\mathbf { L}_{1}}=\left({\mathbf {
L}_{2}}\right)^{\mathbf{T}},\quad {\mathbf {
L}_{4}}=\left({\mathbf { L}_{3}}\right)^{\mathbf{T}}.
\]

\begin{figure}
\begin{center}
\leavevmode \epsfysize=4.5cm \epsfbox{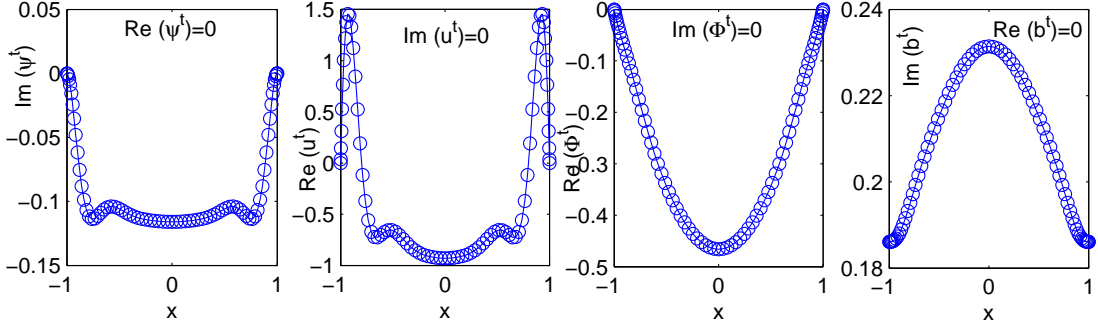}
\end{center}
\caption{{\small (Color online) The adjoint solution
$\mathbb{V}^\dagger$ for ${\cal C} = 0.08$, ${\cal P}_m = 0.001$,
$q = 3/2$ with $R_{_m} \sim 4.9$ and $Q \sim 0.75$.  The open
circles are the values on the Chebyshev grid while the solid lines
are fits.  }} \label{adjoint_plots}
\end{figure}

The adjoint solution ${\mathbf{V}^{\dagger}} \equiv
\mathbb{V}^{\dagger} e^{iQz} + {\rm c.c.}$, in which $
\mathbb{V}^\dagger \equiv
(\Psi^\dagger,u^\dagger,\Phi^\dagger,b^\dagger)^{\mathbf{T}}$ is
such that it satisfies the boundary conditions \beq \Psi^\dagger
=\partial_x\Psi^\dagger = u^\dagger =\Phi^\dagger = \partial_x
b^\dagger = 0, \qquad {\rm at} \ \  x = \pm 1, \eeq in addition to
periodicity in the vertical direction.  In Figure
(\ref{adjoint_plots}), we display an example of
$\mathbb{V}^\dagger$.  We note that it is also an even function
with respect to $x$. The solvability condition
(\ref{order_epsilon2_solvability}) is automatically satisfied on
account of the choice of $Q$ and $R_m$, as discussed in the main
text. To complete this exposition, we explicitly write out the
nonlinear terms appearing in $\mathbf{N_2}$,
\[
\mathbf{N}_2 = A^2 \mathbf{N}_{22} e^{i2Qz} + |A|^2\mathbf{N}_{20}
+ {\rm c.c.,}
\]
where $\mathbf{N}_{22} \equiv
(N^{(\Psi)}_{_{22}},N^{(\Psi)}_{_{22}},
N^{(\Psi)}_{_{22}},N^{(\Psi)}_{_{22}})^{\mathbf{T}}$,
$\mathbf{N}_{20} \equiv (N^{(\Psi)}_{_{20}},N^{(\Psi)}_{_{20}},
N^{(\Psi)}_{_{20}},N^{(\Psi)}_{_{20}})^{\mathbf{T}}$ and \beqa
N^{(\Psi)}_{_{22}} &=& (iQ\Psi_{_{11}})\cdot
D_x(D_x^2-Q^2)\Psi_{_{11}}
-(D_x\Psi_{_{11}})\cdot iQ(D_x^2-Q^2)\Psi_{_{11}} \nonumber \\
& & -{\cal C}\left[ (iQ\Phi_{_{11}})\cdot
D_x(D_x^2-Q^2)\Phi_{_{11}}
-(D_x\Phi_{_{11}})\cdot iQ(D_x^2-Q^2)\Phi_{_{11}}\right] \nonumber \\
N^{(u)}_{_{22}} &=& (iQ\Psi_{_{11}})\cdot D_x u_{_{11}}
-(D_x\Psi_{_{11}})\cdot iQ u_{_{11}} -{\cal C}\left[
(iQ\Phi_{_{11}})\cdot D_x b_{_{11}}
-(D_x\Phi_{_{11}})\cdot iQ b_{_{11}}\right] \nonumber \\
N^{(\Phi)}_{_{22}} & = & -\left[ (iQ\Phi_{_{11}})\cdot D_x
\Psi_{_{11}}
-(D_x\Phi_{_{11}})\cdot iQ \Psi_{_{11}}\right] \nonumber \\
N^{(b)}_{_{22}} &=& (iQ\Psi_{_{11}})\cdot D_x b_{_{11}}
-(D_x\Psi_{_{11}})\cdot iQ b_{_{11}}  -{\cal C}\left[
(iQ\Phi_{_{11}})\cdot D_x u_{_{11}} -(D_x\Phi_{_{11}})\cdot iQ
u_{_{11}}\right], \eeqa and \beqa N^{(\Psi)}_{_{20}} &=&
(iQ\Psi_{_{11}})\cdot (D_x(D_x^2-Q^2)\Psi_{_{11}})^*
-(D_x\Psi_{_{11}})\cdot (iQ(D_x^2-Q^2)\Psi_{_{11}})^* \nonumber \\
& & -{\cal C}\left[ (iQ\Phi_{_{11}})\cdot
(D_x(D_x^2-Q^2)\Phi_{_{11}})^*
-(D_x\Phi_{_{11}})\cdot (iQ(D_x^2-Q^2)\Phi_{_{11}})^*\right] \nonumber \\
N^{(u)}_{_{20}} &=& \underline{ (iQ\Psi_{_{11}})\cdot (D_x
u_{_{11}})^*} -(D_x\Psi_{_{11}})\cdot (iQ u_{_{11}})^* -\left[
(iQ\Phi_{_{11}})\cdot (D_x b_{_{11}})^*
-(D_x\Phi_{_{11}})\cdot (iQ b_{_{11}})^*\right] \nonumber \\
N^{(\Phi)}_{_{20}} & = & -\left[ (iQ\Phi_{_{11}})\cdot (D_x
\Psi_{_{11}})^*
-(D_x\Phi_{_{11}})\cdot (iQ \Psi_{_{11}})^*\right] \nonumber \\
N^{(b)}_{_{20}} &=& (iQ\Psi_{_{11}})\cdot (D_x b_{_{11}})^*
-(D_x\Psi_{_{11}})\cdot (iQ b_{_{11}})^*  -\left[
(iQ\Phi_{_{11}})\cdot (D_x u_{_{11}})^*
-(D_x\Phi_{_{11}})\cdot (iQ u_{_{11}})^*\right].\nonumber \\
&& \label{N20_expressions} \eeqa Superscript``*" on any given
quantity (i.e. $f^*$) denotes the complex conjugation of the said
quantity.  We solve for the quantities
$\mathbb{V}_{_{22}},\mathbb{V}_{_{21}},\mathbb{V}_{_{20}}$ using
the Chebyshev collocation technique developed for the linear
theory and show an example of these results in Figure
\ref{order_epsilon_2_functions}. We note that the functions of
$\mathbb{V}_{_{22}},\mathbb{V}_{_{20}}$ are odd with respect to
$x$ while those of $\mathbb{V}_{_{21}}$ are even with respect to
$x$.  We also call attention to the fact that since $N^{(b)}_{20}
+ \left(N^{(b)}_{20}\right)^* = 0$, $b_{_{20}}$ is also zero. This
is significant because it is really a consequence of a second
condition that must be met to ensure the existence of
 a solution at this order.  In particular, inspection of the $z$ independent
component of the equation describing the evolution of $b_2$, i.e.,
\beq R_{_m}^{-1}\partial_x^2 b_{_{20}} = |A|^2\left[N^{(b)}_{20} +
\left(N^{(b)}_{20}\right)^*\right], \label{b20_equation} \eeq
shows that in order for there to be a solution to $b_{_{20}}$,
which satisfies the boundary conditions $\partial_x b_{_{20}} = 0$
at $x=\pm 1$, a condition must be met with respect to the terms on
the right-hand side of (\ref{b20_equation}). This criterion is
most simply seen by (i) integrating this equation from $x=-1$ to
$x=1$, (ii) applying boundary conditions, (iii) leaving the
requirement
\[
0=|A|^2\int_{-1}^1\left[N^{(b)}_{20} +
\left(N^{(b)}_{20}\right)^*\right]dx.
\]
However, this relationship is automatically satisfied on account
of the fact that $Re(N^{(b)}_{20}) = 0$, as noted above.
\begin{figure}
\begin{center}
\leavevmode \epsfysize=8.5cm
\epsfbox{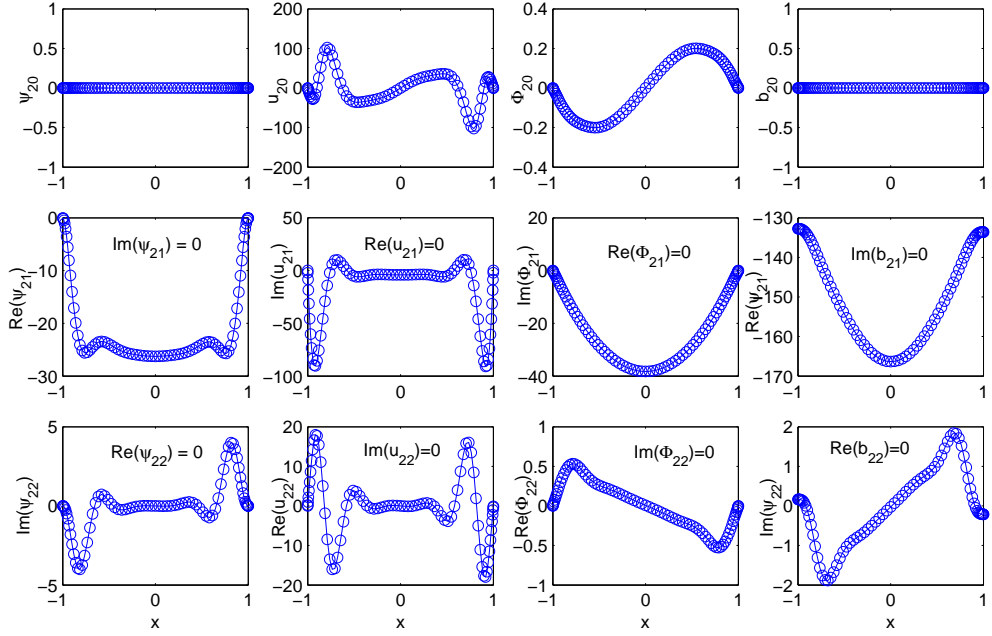}
\end{center}
\caption{{\small (Color online) The second order solutions
$\mathbb{V}_{22}, \mathbb{V}_{21}, \mathbb{V}_{20}$ for ${\cal C}
= 0.08$, ${\cal P}_m = 0.001$, $q = 3/2$ with $R_{_m} \sim 4.9$
and $Q \sim 0.75$.  Note the pronounced boundary layers in
$u_{22},u_{21}$, $u_{20}$ and, to a lesser degree, $\Psi_{21}$.
}} \label{order_epsilon_2_functions}
\end{figure}
{We complete the solution to this order by explicitly writing out
the form for $\tilde u_{20}$ which is associated with $\partial_Z
B$ - c.f. (\ref{order_2_general_solution}).  The equation
governing its structure is \beq {\cal R}^{-1}\partial_x^2 \tilde
u_{20} = -{\cal C}, \eeq yielding the solution \beq \tilde u_{20}
= \frac{1}{2}{\cal C}{\cal R} (x^2-1). \label{tilde_u20_solution}
\eeq}
\par
Finally, at order $\epsilon^3$, the equations are \beq
{\mathbf{{\cal D}}}\partial_{_T}{\mathbf{V}_{_1}}
+{\mathbf{N}_{3}} = {\cal L}{\mathbf{V}_{_3}}
+\mathbf{\tilde{{\cal L}}_1}\partial_{_Z}{\mathbf{V}_{_2}}
+\mathbf{\tilde{{\cal L}}_2}\partial^2_{_Z}{\mathbf{V}_{_1}}
+\mathbf{\tilde{{\cal G}}}{\mathbf{V}_{_1}}.
\label{order_epsilon3_equation} \eeq The terms of the nonlinear
functional ${\mathbf{N}_{3}}$ have the more detailed following
expansion \beq {\mathbf{N}_{3}} = A^3 \mathbf{N}_{33}e^{i3Qz} +A
\partial_{_Z}A \mathbf{N}_{32}e^{i2Qz} + A|A|^2
\mathbf{N}_{31}e^{iQz} + A \partial_{_Z}B \mathbf{\tilde
N}_{31}e^{iQz} + A^*\partial_{_Z}A \mathbf{N}_{30} + {\rm c.c.}
\label{N_3_expansion} \eeq

In order for there to be a solution at this order, the same
solvability condition discussed earlier must also be satisfied for
this equation.  Those terms in the above expression subject to the
solvability condition are the ones resonant with $e^{iQz}$ (see
the definition of $\mathbf{V}^\dagger$ above).  Our goal in this
work is to satisfy this solvability condition, at this order.  In
turn, this means that the only two terms from the nonlinear
expression that will concern us here with this {\em first
solvability} (see below) will be the expressions involving
$\mathbf{N}_{31}$ and $\mathbf{\tilde N}_{31}$.  The explicit
forms for these expressions are given in Appendix C (in order not
to clutter this exposition).  The main feature to note about these
functions is that $\mathbf{N}_{31}$ is even with respect to $x$
while $\mathbf{\tilde N}_{31}$ is odd.  The ramifications of these
facts are explained below.  The solvability condition means taking
the inner product of (\ref{order_epsilon3_equation}) with
$\mathbf{V}^\dagger$, revealing \beq a\partial_{_T} A + c A|A|^2 +
\tilde c A \partial_{_Z} B = b A + h \partial^2_{_Z} A, \eeq where
\beqa a \equiv \left<\mathbb{V}^\dagger \cdot {\cal D}
\mathbb{V}_{_{11}}^{*}\right>, \quad c&\equiv&
\left<\mathbb{V}^\dagger \cdot \mathbf{N}_{31}^{*}\right>, \quad
\tilde c\equiv \left<\mathbb{V}^\dagger \cdot \mathbf{\tilde
N}_{31}^{*}\right>, \quad b \equiv \left<\mathbb{V}^\dagger \cdot
({\cal G} \mathbb{V}_{_{11}})^{*}\right>, \nonumber\\ \quad h
&\equiv& \left<\mathbb{V}^\dagger \cdot \left(\mathbf{\tilde{{\cal
L}}_1} \mathbb{V}_{_{21}} +\mathbf{\tilde{{\cal
L}}_2}\mathbb{V}_{_{11}}\right)^{*} \right>. \label{Acoeff_list}
\eeqa Inspecting the expressions comprising $\mathbf{\tilde
N}_{31}$ reveals that they are odd symmetric with respect to $x$.
This result means that the expression $\tilde c = 0$ and,
consequently, it means that the evolution equation for the field
$A$ evolves according to \beq
\partial_{_T} A = \lambda A + D \partial^2_{_Z} A - \alpha A|A|^2,
\qquad {\rm where} \quad \lambda \equiv b/a,\quad D \equiv
h/a,\quad \alpha = c/a, \eeq independent of the second field
quantity $B$.  \emph{This decoupling is a direct consequence of
the symmetries preserved in the thin gap Taylor-Couette limit.
This will not be the case in the general-gap magnetized
Taylor-Couette problem.}
\begin{figure}
\begin{center}
\leavevmode \epsfysize=4.5cm \epsfbox{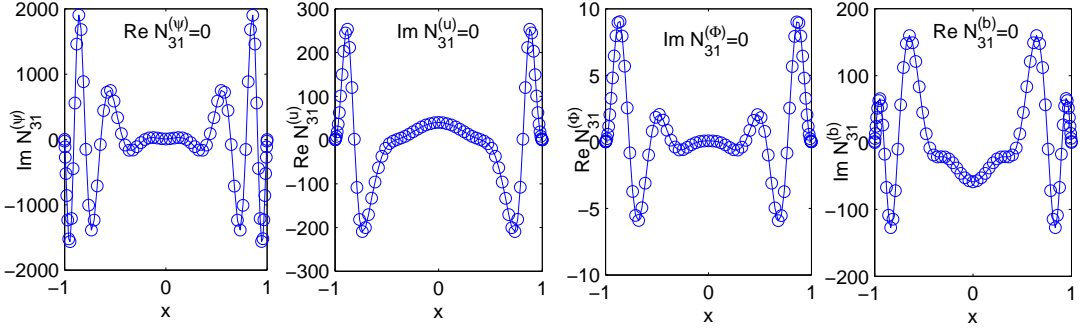}
\end{center}
\caption{{\small (Color online) The non-linear function
$\mathbf{N}_{31}$ for ${\cal C} = 0.08$, ${\cal P}_m = 0.001$, $q
= 3/2$ with $R_{_m} \sim 4.9$ and $Q \sim 0.75$.  }}
\label{n31_plots}
\end{figure}

As noted earlier, there is a second solvability condition, which
must be enforced on the order $\epsilon^3$ $B$-equation.  This
becomes necessary on the component of the solution for which there
is no explicit $z$ dependence.  More explicitly, writing out that
component of the equation: \beq {R}_{_m}^{-1}\partial_x^2 b_{30} =
\partial_{_T} B - { R}_{_m}^{-1}\partial^2_{_Z} B - \tilde u_{20}
\partial_{_Z}^2 B + \Bigl[A^*\partial_{_Z}A N^{(b)}_{30} + {\rm
c.c.}\Bigr]. \label{b30_equation} \eeq {The solvability condition
can be readily inferred by integrating (\ref{b30_equation}) from
$x=-1$ to $x=1$ and requiring that $\partial_x b_{30} = 0$ at
$x=\pm 1$.  This procedure is equivalent to taking the inner
product of (\ref{order_epsilon3_equation}) multiplied by
$\mathbb{U}_{_{11}}$, thus revealing the second envelope equation
\[
\partial_{_T} B = \left(\frac{1}{R_{_m}}+\frac{{\cal C}{\cal R}}{3}\right) \partial^2_{_Z} B - p \partial_{_Z}|A^2|,
\]
where $p \equiv \left<Re\left(N^{(b)}_{30}\right)\right>$.
However, the odd symmetry property of $N^{(b)}_{30}$ means that
$p=0$. Thus, the second solvability condition yields the simple
diffusion equation, \beq
\partial_{_T} B =  \left(\frac{1}{R_{_m}}+\frac{{\cal C}{\cal R}}{3}\right)  \partial^2_{_Z} B.
\eeq}
\par

\section{The terms for $\mathbf{N}_{31}$}
We define for notational convenience the symbols $\nabla_1^2
\equiv D_x^2 -Q^2$ and $\nabla_2^2 \equiv D_x^2 - 4Q^2$.  It
follows that the expressions are \beqa N^{(\Psi)}_{_{31}} &=&
(i2Q\Psi_{_{22}})(D_x\nabla_1^2\Psi_{_{11}})^* -\left[
(D_x\Psi_{_{22}})\cdot(iQ\nabla_1^2\Psi_{_{11}})^*
+(D_x\Psi_{_{20}})\cdot(iQ\nabla_1^2\Psi_{_{11}})
\right] \nonumber \\
& & +(iQ\Psi_{_{11}})^*\cdot(D_x\nabla_2^2\Psi_{_{22}})
+(iQ\Psi_{_{11}})\cdot D_x D_x^2\Psi_{_{20}}
-(D_x\Psi_{_{11}})^*\cdot(i2Q\nabla_2^2\Psi_{_{22}})
\nonumber \\
& & -{\cal C}\left\{(i2Q\Phi_{_{22}})(D_x\nabla_1^2\Phi_{_{11}})^*
-\left[ (D_x\Phi_{_{22}})\cdot(iQ\nabla_1^2\Phi_{_{11}})^*
+(D_x\Phi_{_{20}})\cdot(iQ\nabla_1^2\Phi_{_{11}}) \right] \right\}
\nonumber \\
& & -{\cal
C}\left\{(iQ\Phi_{_{11}})^*\cdot(D_x\nabla_2^2\Phi_{_{22}})
+(iQ\Phi_{_{11}})\cdot D_x D_x^2\Phi_{_{20}}
-(D_x\Phi_{_{11}})^*\cdot(i2Q\nabla_2^2\Phi_{_{22}})
\right\}, \\
N^{(u)}_{_{31}} &=& (i2Q\Psi_{_{22}})(D_x u_{_{11}})^* -\left[
(D_x\Psi_{_{22}})\cdot(iQu_{_{11}})^*
+(D_x\Psi_{_{20}})\cdot(iQu_{_{11}})
\right] \nonumber \\
& & +(iQ\Psi_{_{11}})^*\cdot(D_x u_{_{22}})
+\underline{(iQ\Psi_{_{11}})\cdot D_x u_{_{20}}}
-(D_x\Psi_{_{11}})^*\cdot(i2Qu_{_{22}})
\nonumber \\
& & -{\cal C}\left\{(i2Q\Phi_{_{22}})(D_x b_{_{11}})^* -\left[
(D_x\Phi_{_{22}})\cdot(iQb_{_{11}})^*
+(D_x\Phi_{_{20}})\cdot(iQb_{_{11}}) \right] \right\}
\nonumber \\
& & -{\cal C}\left\{(iQ\Phi_{_{11}})^*\cdot(D_x b_{_{22}})
+(iQ\Phi_{_{11}})\cdot D_x b_{_{20}} -(D_x
\Phi_{_{11}})^*\cdot(i2Qb_{_{22}})
\right\}, \label{N31_u_expression}\\
N^{(\Phi)}_{_{31}} &=& -\left\{ (i2Q\Psi_{_{22}})(D_x
\Psi_{_{11}})^* -\left[ (D_x \Psi_{_{22}})\cdot(iQ\Psi_{_{11}})^*
+(D_x \Psi_{_{20}})\cdot(iQu\Psi_{_{11}})
\right]\right\} \nonumber \\
& & \left\{(iQ\Psi_{_{11}})^*\cdot(D_x \Psi_{_{22}})
+(iQ\Psi_{_{11}})\cdot D_x \Psi_{_{20}}
-(D_x \Psi_{_{11}})^*\cdot(i2Q\Psi_{_{22}})\right\},\\
N^{(b)}_{_{31}} &=& (i2Q\Psi_{_{22}})(D_x b_{_{11}})^* -\left[
(D_x\Psi_{_{22}})\cdot(iQb_{_{11}})^*
+(D_x\Psi_{_{20}})\cdot(iQb_{_{11}})
\right] \nonumber \\
& & +(iQ\Psi_{_{11}})^*\cdot(D_x b_{_{22}}) +(iQ\Psi_{_{11}})\cdot
D_x b_{_{20}} -(D_x\Psi_{_{11}})^*\cdot(i2Qb_{_{22}})
\nonumber \\
& & -\left\{(i2Q\Phi_{_{22}})(D_x u_{_{11}})^* -\left[
(D_x\Phi_{_{22}})\cdot(iQu_{_{11}})^*
+(D_x\Phi_{_{20}})\cdot(iQu_{_{11}}) \right] \right\}
\nonumber \\
& & -\left\{(iQ\Phi_{_{11}})^*\cdot(D_x u_{_{22}})
+\underline{(iQ\Phi_{_{11}})\cdot D_x u_{_{20}}} -(D_x
\Phi_{_{11}})^*\cdot(i2Qu_{_{22}}) \right\} \eeqa and \beqa
\tilde N^{(\Psi)}_{_{31}} &=& 0 \\
\tilde N^{(u)}_{_{31}} &=& (iQ\Psi_{_{11}})\cdot D_x\tilde u_{_{20}} \\
\tilde N^{(\Phi)}_{_{31}} &=& 0 \\
\tilde N^{(b)}_{_{31}} &=& -(iQ\Phi_{_{11}})\cdot D_x\tilde
u_{_{20}} \eeqa

\section{On the $\order{{\cal P}_m^{-4/3}}$ dependence of $\alpha$.}\label{alpha_dependence}
There are a number of terms comprising the integral expression
leading to the quantity $\alpha$.  Of these, there are a few that
dominate its expression when ${\cal P}_m$ is small.  In the
following, we will sketch out one way to understand the
$\order{{\cal P}_m^{-4/3}}$ scaling behavior of $\alpha$. (Note
that because we consider the behavior of this system by holding
${\cal R}_m = {\rm R}_m$ fixed, we will speak about the general
scaling dependencies of quantities on ${\cal P}_m$ and ${\cal
R}^{-1}$ interchangeably as they are, in effect, equivalent under
this constraint.)
\par
We noted earlier that the lowest order functions comprising
$\mathbb{V}_{_{11}}$ exhibit boundary layers of spatial extent
$\order{{\cal P}_m^{1/3}}$ for ${\cal P}_m \ll 1$. Especially
acute in this respect is the function for the azimuthal velocity
perturbation, $u_{_{11}}$. At the next order, we find that the
equation for $u_{_{20}}$ (i.e. the second order azimuthal velocity
function with no dependence on the $z$ coordinate) is simply \beq
{\cal R}^{^{-1}}\partial_x^2 u_{_{20}} = N^{(u)}_{_{20}}.
\label{u20_eqn} \eeq Inspection of (\ref{N20_expressions}) shows
that $N^{(u)}_{_{20}}$ is dominated by the underlined term
containing the expression $D_x u_{_{11}}$.  (Note that it is true
that the other quantities also, in principle, have boundary layers
as well but the one associated with $D_x u_{_{11}}$ dominates -
inspection of Fig.~\ref{marginal_eigenfunctions} readily shows.)
This means that in the boundary layer regions, $N^{(u)}_{_{20}}$
scales as $\order{{\cal P}_m^{-1/3}}$ while in the interior it
remains $\order 1$.  It follows from inspection of (\ref{u20_eqn})
that $u_{_{20}}$ scales as $\order{{\cal P}_m^{-4/3}}$ in the
boundary layers and $\order{{\cal P}_m^{-1}}$ in the bulk
interior.\par Now we turn to an inspection of the expression
leading to $\alpha$, namely term $c$ of (\ref{Acoeff_list}) which
is composed, in part, of the integrals over the domain of the
products $u^\dagger\cdot N^{(u)}_{_{31}}$ and $b^\dagger\cdot
N^{(b)}_{_{31}}$.  Since $u^\dagger$ and $b^\dagger$ remain
$\order 1$ over the entirety of the domain, it remains for us to
evaluate the behavior of $N^{(u)}_{_{31}}$ and $N^{(b)}_{_{31}}$
over the domain.  Although there are several terms that
contribute, one of the terms is most dominant: the underlined
expressions in (\ref{N31_u_expression}), which involves $D_x
u_{_{20}}$.  Because there is a derivative, the scale of $D_x
u_{_{20}}$ gets amplified by another factor of ${\cal P}_m^{-1/3}$
in the boundary layer region. Given what we have established thus
far about the character and profile of $u_{_{20}}$, it follows
that $D_x u_{_{20}}$ is $\order{{\cal P}_m^{-5/3}}$ in the
boundary layer regions while it is $\order{{\cal P}_m^{-1}}$ in
the interior. It means, therefore, that the profiles of
$N^{(u)}_{_{31}}$ and $N^{(b)}_{_{31}}$ similarly reflect this
character on the $x$ domain.\par Thus, for instance, to determine
the order of magnitude of the integral of $b^\dagger
N^{(b)}_{_{31}}$ over the domain, we should break up the integral
into parts separating out the interior region and boundary layers.
Writing $\delta = {\cal P}_m^{1/3}$ we have
\[
\int_{-1}^{1}b^\dagger N^{(b)}_{_{31}}
 dx = \underbrace{\int_{-1}^{-1+\delta}b^\dagger N^{(b)}_{_{31}} dx}
_{\order{{\cal P}_m^{-5/3}}\cdot \order{{\cal P}_m^{1/3}}} +
\underbrace{\int_{-1+\delta}^{1-\delta} b^\dagger N^{(b)}_{_{31}}
dx}_ {\order{{\cal P}_m^{-1}}\cdot \order{1}} +
\underbrace{\int_{1-\delta}^{1} b^\dagger N^{(b)}_{_{31}} dx}_
{\order{{\cal P}_m^{-5/3}}\cdot \order{{\cal P}_m^{1/3}}}.
\]
In other words, because the length scale of the interior zone is
$\order 1$ and the scale of $b^\dagger N^{(b)}_{_{31}} =
\order{{\cal P}_m^{-1}}$ in that region, the contribution to the
integral from this part is $\order{{\cal P}_m^{-1}}$.  On the
other hand, because the length scale of the boundary layer(s) is
$\order{{\cal P}_m^{1/3}}$, while the value of $b^\dagger
N^{(b)}_{_{31}}$ scales as $\order{{\cal P}_m^{-5/3}}$ in those
regions, it follows that the contribution to the total integral
from these zones is $\order{{\cal P}_m^{-4/3}}$. The same
reasoning follows for the integral of $u^\dagger N^{(u)}_{_{31}}$
over the domain.  \par The nonlinear readjustment occurring in the
boundary layers dominates the scale of $\alpha$ and we can
conclude that the dominant process leading to saturation occurs in
the boundary layers.  We note also that had there been no boundary
layers, then $\alpha$ would scale as $\order{{\cal P}_m^{-1}}$
because of the scale of $u_{_{20}}$, which is \emph{always at
least} $\order{{\cal P}_m^{-1}}$ on account of (\ref{u20_eqn}).
This directly relates to the problem investigated in UMR06, in
which boundary layers are suppressed on account of the boundary
conditions employed in that study.  In that case, the saturation
process gets contributions from the entirety of the domain and not
just the boundary layers.


\begin{thebibliography}{}
\bibitem{note_1}  Note that the
quantity $c$ (B40),
which is the cubic term of the real GLE (\ref{landau_eqn}),
is primarily the weighted measure of this delivered torque in the limit
where ${\cal P}_m \ll 1$.
\bibitem{note_2} This inner product, symbolized by the bracket operation $<\mathbb F> \equiv
\frac{Q}{2\pi}\int_{_{-1}}^{1}\int_{-\pi/Q}^{\pi/Q}{\mathbb F}(x,z)\ dxdz$, will be
invoked whenever a solvability condition is required.
\end{thebibliography}
\end{document}